\documentclass[reprint,amsmath,amssymb,aps,prl,floatfix,superscriptaddress]{revtex4-1}

\usepackage[T1]{fontenc}
\usepackage[utf8]{inputenc}
\usepackage[english]{babel}
\usepackage{graphicx}
\usepackage[colorlinks=true,urlcolor=blue,linkcolor=blue,
            citecolor=blue]{hyperref}
\usepackage{times}
\usepackage{bm}
\usepackage{placeins}

\newcommand{\qvec}{{\bf q}}
\newcommand{\pvec}{{\bf p}}
\newcommand{\rvec}{{\bf r}}
\newcommand{\Rvec}{{\bf R}}
\newcommand{\kvec}{{\bf k}}
\newcommand{\Gvec}{{\bf G}}

\newcommand{\bibtitle}[1]{\textit{#1},}

\begin{document}
\title{Correlation energies of the high-density spin-polarized
  electron gas to meV accuracy}

\author{Michele Ruggeri}
  \affiliation{Max Planck Institute for Solid State Research,
               Heisenbergstr.\ 1, 70569 Stuttgart, Germany}
\author{Pablo L\'opez R\'ios}
  \affiliation{Max Planck Institute for Solid State Research,
               Heisenbergstr.\ 1, 70569 Stuttgart, Germany}
  \affiliation{Theory of Condensed Matter Group, Cavendish Laboratory,
               J. J. Thomson Avenue, Cambridge CB3 0HE, UK}
\author{Ali Alavi}
  \affiliation{Max Planck Institute for Solid State Research,
               Heisenbergstr.\ 1, 70569 Stuttgart, Germany}
  \affiliation{University Chemical Laboratory,
               Lensfield Road, Cambridge CB2 1EW, UK}

\begin{abstract}
We present a novel combination of quantum Monte Carlo methods and a
finite size extrapolation framework with which we calculate the
thermodynamic limit of the exact correlation energy of the polarized
electron gas at high densities to meV accuracy, $-40.44(5)$ and
$-31.70(4)$ mHa at  $r_{\rm s}=0.5$ and $1$, respectively.
The fixed-node error is characterized and found to exceed $1$ mHa, and
we show that the magnitude of the correlation energy of the polarized
electron gas is underestimated by up to $6$ meV by the Perdew-Wang
parametrization, for which we suggest improvements.
\end{abstract}

\maketitle


The uniform (or homogeneous) electron gas (UEG) \cite{loos_ueg_2016}
is a system consisting of electrons in a neutralizing uniform
background intended to model the behavior of electrons in metals
\cite{huotari_ueg_2010}.
This system is of crucial importance in understanding the nature of
electronic correlation, and is of huge practical relevance since
knowledge of the correlation energy of the UEG as a function of its
homogeneous density can be used as a key ingredient in the description
of the behavior of electrons in real systems \cite{wigner_ecorr_1938,
hohemberg_DFT_1956, kohn_DFT_1964}.

Despite its seeming simplicity, the complex correlations caused by the
long-ranged character of the Coulomb interaction require the use of
explicit many-body methods to accurately characterize the UEG.
The release-node diffusion Monte Carlo calculations of Ceperley and
Alder (CA) \cite{ceperley_1980} provided data connecting the analytic
high-density \cite{gellmann_1957, hoffman_1992} and low-density
\cite{carr_1961} limits of the correlation energy, and enabled the
development of parametrizations over the entire density range
\cite{vosko_ecorr_1980, perdew_lda_1981, perdew_wang_1992} which are
routinely used in density functional theory calculations.

\begin{figure}[!hbt]
  \begin{center}
    \includegraphics[width=\columnwidth]{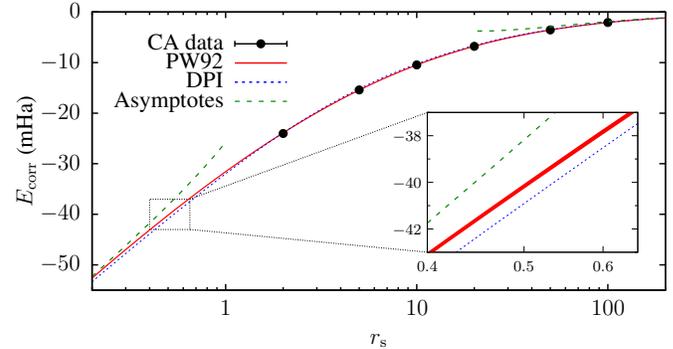}
    \caption{
      Correlation energy of the polarized UEG as a function of
      $r_{\rm s}$.
      Shown are the CA data \cite{ceperley_1980}, the PW92
      parametrization \cite{perdew_wang_1992}, and the DPI
      parametrization \cite{sun_dpi_2010, loos_dpi_2011,
      bhattarai_dpi_2018}.
      The inset magnifies the region around $r_{\rm s}=0.5$.
      The width of the PW92 curve represents its statistical
      uncertainty.}
    \label{fig:pw_ecorr}
  \end{center}
\end{figure}

The Perdew-Wang parametrization of the correlation energy of the UEG
(PW92) \cite{perdew_wang_1992} has become a cornerstone in the
construction of density functionals over the past three decades.
The PW92 form contains five parameters, of which two are determined
from analytic high-density constraints and three by fitting to the CA
data.
More recently, a ``density parameter interpolation'' (DPI)
parametrization was proposed \cite{sun_dpi_2010, loos_dpi_2011,
bhattarai_dpi_2018} that is constructed by imposing four high-density
and three low-density constraints on a seven-parameter functional
form, thus requiring (almost) no quantum Monte Carlo input.
In Fig.\ \ref{fig:pw_ecorr} we plot the PW92 and DPI parametrizations
for the fully-polarized electron gas, along with the asymptotes
defined in Refs.\ \onlinecite{sun_dpi_2010, loos_dpi_2011,
bhattarai_dpi_2018}, as a function of $r_{\rm s}$, the radius of the
sphere containing one electron on average divided by the Bohr radius.
While the two parametrizations are in excellent agreement at low
densities, they differ by $\sim 20$ meV at densities relevant to
systems with all-electron nuclei \cite{zupan_1998} and solids at high
pressures.
The cumulative effect of incurring these small errors in the
parametrized correlation energy could result in a significant bias in
computed properties, including total and relative energy estimates.

In this Letter we use a combination of full configuration-interaction
quantum Monte Carlo (FCIQMC) and fixed-node diffusion Monte Carlo
(DMC) to compute the correlation energy of the fully spin-polarized
three-dimensional UEG at $r_{\rm s}=0.5$ and $1$ to meV accuracy.
Building upon existing knowledge of finite size errors in DMC
\cite{lin_ta_2001, chiesa_fse_2006, drummond_fse_2008,
holzmann_fse_2016}, we propose an extrapolation procedure which we
find to be much more accurate than previous approaches.
By extrapolating the fixed node energy and the fixed node error to the
thermodynamic limit we obtain the exact correlation energies at
$r_{\rm s}=0.5$ and $1$.
We are thus able to resolve the discrepancy between the values of the
PW92 and DPI parametrizations at high densities, and we discuss ways
to improve their accuracy.


We simulate finite systems of $N$ same-spin electrons in a cubic
simulation cell at fixed homogeneous densities using DMC and FCIQMC.
Note that we report energies per electron and use Hartree atomic units
($\hbar = m_e = |e| = 4\pi\epsilon_0 = 1$) throughout.
Full details about the methodology and calculations are given in
the Supplemental Material \cite{supplemental}.

The variational Monte Carlo (VMC) \cite{mcmillan_1965,
toulouse_emin_2007, umrigar_emin_2007} and fixed-node DMC methods
\cite{ceperley_1980, foulkes_rmp_2001, umrigar_dmc_1993, zen_2016}
have been extensively used to study the UEG \cite{kwon_backflow_1998,
holzmann_backflow_2003, lopezrios_backflow_2006, gurtubay_ueg_2010,
spink_ueg_2013} using Slater-Jastrow trial wave functions, formed by
the Hartree-Fock (HF) determinant multiplied by a Jastrow correlation
factor \cite{drummond_jastrow_2004, lopezrios_jastrow_2012}, often in
combination with backflow transformations
\cite{pandharipande_backflow_1973, schmidt_backflow_1979,
kwon_backflow_1998, holzmann_backflow_2003, lopezrios_backflow_2006}.
While these wave functions are reasonably sophisticated, the energy
obtained by the DMC method incurs a positive bias, referred to as the
fixed-node error $\varepsilon_{\rm FN}$, caused by the restrictions
imposed by the fixed-node approximation \cite{anderson_1976,
ceperley_1980}.

The FCIQMC method explictly operates in the basis of antisymmetric
Slater determinants, thus avoiding the need for a fixed node
approximation \cite{booth_fciqmc_2009}.
The initiator approximation \cite{cleland_initiator_2010} allows the
efficient exploration of this vast Hilbert space, and has enabled the
successful application of FCIQMC to systems of interest in quantum
chemistry and condensed matter physics \cite{booth_ionization_2010,
cleland_affinities_2011, booth_c2_2011, cleland_firstrow_2012,
schwarz_2015, kersten_2016}, including the unpolarized UEG
\cite{shepherd_fciqmc_2012, shepherd_ueg_2012, neufeld_ueg_2017,
luo_2018}.
FCIQMC calculations use finite basis sets, and the infinite basis set
limit can be estimated by extrapolation, as is standard practice in
quantum chemistry \cite{helgaker_mol_es_2014}.
We find that the basis set error for the polarized UEG is well
described by a quadratic function of the inverse basis set size
\cite{supplemental}, in contrast with the linear dependence found for
the unpolarized UEG \cite{shepherd_fciqmc_2012}.

\begin{figure}[!hbt]
  \begin{center}
    \includegraphics[width=\columnwidth]{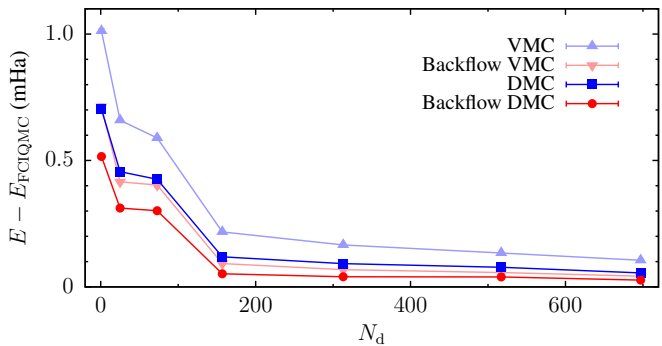}
    \caption{
      VMC and DMC energies of the polarized 19-electron gas at $r_{\rm
      s} = 1$ (at $\Gamma$) relative to FCIQMC, as a function of the
      number of determinants in the wave function, both without and
      with backflow transformations.}
    \label{fig:mdet}
  \end{center}
\end{figure}

We assess the quality of our FCIQMC energies by comparison with VMC
and DMC energies for increasingly accurate trial wave functions.
We construct multi-determinantal wave functions for the 19-electron
gas at $r_{\rm s}=1$ by truncating the FCIQMC wave function to the
$N_{\rm d}$ leading determinants, with symmetry-equivalent
determinants grouped together.
The results, obtained using the \textsc{casino} code \cite{casino},
are plotted in Fig.\ \ref{fig:mdet} against $N_{\rm d}$.

The variational convergence of our VMC and DMC energies towards the
FCIQMC energy is consistent with FCIQMC being exact for this system.
The best backflow DMC energy is only $0.027(5)$ mHa higher than the
FCIQMC energy, and is to our knowledge the most accurate DMC energy
for this system reported to date.


The finite size error in the energy of the UEG consists of a
contribution which varies smoothly with $N$ and a quasirandom
contribution, which must be eliminated to enable a clean extrapolation
of the smooth part.
Twist averaging \cite{lin_ta_2001} substantially reduces quasirandom
fluctuations by averaging over wave vector offsets in the Brilluoin
zone.
In DMC we sample the Brillouin zone randomly, while for our FCIQMC
calculations we divide the Brillouin zone into regions of equal total
momentum and run FCIQMC calculations in each of these regions
\cite{shepherd_thesis_2013}, which we are able to characterize exactly
\cite{supplemental}.
In selected cases we perform the basis-set extrapolation in one region
and use the extrapolation parameters for the others, which reduces the
number of required FCIQMC calculations considerably
\cite{supplemental}.
In what follows we discuss twist averaged energies except when stated
otherwise.

Quasirandom errors are further reduced by subtracting the finite size
error in the HF kinetic energy $\Delta K(N) = K(N)-K(\infty)$ from the
DMC total energy \cite{ceperley_1980, drummond_fse_2008}.
Additionally, we find that the residual quasirandom fluctuations are
highly correlated with those in the HF exchange energy $X(N)$.
The exchange energy is a particularly slowly varying function at large
$N$, so subtracting $X(N)-X(\infty)$ would complicate the
extrapolation.
However, Drummond \textit{et al.\@} \cite{drummond_fse_2008} found
that the leading-order contribution to the finite size error in $X(N)$
for an electron gas is exactly $h_2 N^{-2/3}$, where $h_2=-\frac{3
\epsilon_1}{16\pi} r_{\rm s}^{-1}$ for the polarized UEG and
$\epsilon_1=5.674594959$ for simple cubic simulation cells
\cite{drummond_fse_2008, supplemental}.
We therefore obtain the thermodynamic limit by extrapolation of
$E_{\rm tot}^{\rm FN}(N)-\Delta K(N)-\Delta X(N)$, where $\Delta X(N)
= X(N)-X(\infty) - h_2 N^{-2/3}$.
This is equivalent to extrapolating $E_{\rm corr}^{\rm FN}(N) + h_2
N^{-2/3}$, and in practice we work with the correlation energy
directly.

We model the smooth part of the finite size error as a polynomial in
$N^{-1/3}$, in agreement with the form of the contributions found by
Ref.\ \onlinecite{drummond_fse_2008}, and we find that the use of the
above treatment of quasirandom fluctuations enables the use of fairly
high-order polynomials.
Chiesa \textit{et al.\@} \cite{chiesa_fse_2006} showed that the
leading-order contribution to the finite size error in the total DMC
energy of an electronic system is $t_3 N^{-1}$, where
$t_3=-\frac{\sqrt{3}}2 r_{\rm s}^{-3/2}$ for the polarized UEG.
Since beyond-leading-order contributions to both $\Delta K(N)$ and
$\Delta X(N)$ are proportional to $N^{-4/3}$ \cite{lin_ta_2001,
drummond_fse_2008}, the DMC correlation energy satisfies
\begin{equation}
  \label{eq:dmc_ecorr_tdl}
  \begin{split}
  E_{\rm corr}^{\rm FN}(N) & + h_2 N^{-2/3} - t_3 N^{-1} = \\
                           & c_0 + c_4 N^{-4/3} + c_5 N^{-5/3}
                                 + c_6 N^{-2} + \ldots \;,
  \end{split}
\end{equation}
where $\{c_n\}$ are density-dependent parameters.

\begin{figure}[!hbt]
  \begin{center}
    \includegraphics[width=\columnwidth]{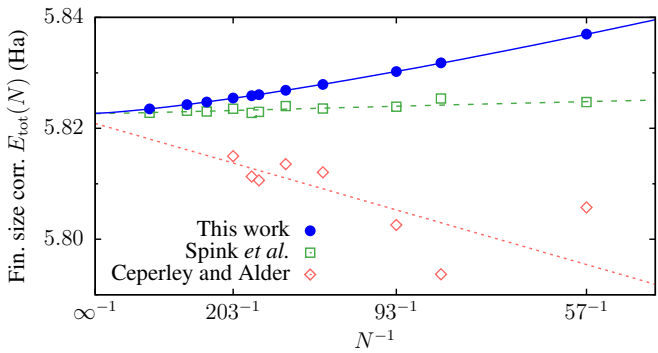}
    \caption{
      Finite size corrected fixed node energies of the polarized UEG
      at $r_{\rm s} = 0.5$ as a function of $N^{-1}$.
      Our results correspond to the solid circles and the solid line.
      Data replicating the finite size treatment of Ceperley and Alder
      \cite{ceperley_1980} (open diamonds and dotted line) and Spink
      \textit{et al.\@} \cite{spink_ueg_2013} (open squares) are also
      plotted, along with an extrapolation of the latter using
      $N^{-4/3}$ and $N^{-5/3}$ terms.}
    \label{fig:tdl_compare}
  \end{center}
\end{figure}

We perform DMC calculations of the polarized UEG using the
Slater-Jastrow wave function at system sizes $15\leq N \leq 515$ at
$r_{\rm s}=0.5$ and $15\leq N \leq 1021$ at $r_{\rm s}=1$, and we use
Eq.\ \ref{eq:dmc_ecorr_tdl} to obtain the thermodynamic limit of the
fixed node correlation energy, setting $h_2$ and $t_3$ to their
analytic values and treating $c_0$, $c_4$, $c_5$, and $c_6$ as fit
parameters.
We do not use backflow or multi-determinants to avoid introducing
wave function optimization noise in our DMC energies.
The magnitude of quasirandom fluctuations has been observed to decay
as $N^{-1}$ \cite{lin_ta_2001}, so we use $N^2$ as weights in our
fits.
In Fig.\ \ref{fig:tdl_compare} we plot $E_{\rm tot}(N) - \Delta K(N) -
\Delta X(N) - t_3 N^{-1}$ and our extrapolation (solid circles and
solid line) at $r_{\rm s} = 0.5$ as a function of $N^{-1}$.
These results numerically confirm the absence of additional
contributions to Eq.\ \ref{eq:dmc_ecorr_tdl} at order $N^{-1}$ or
slower.

In Fig.\ \ref{fig:tdl_compare} we also compare our extrapolation with
other approaches used in the literature.
Ceperley and Alder \cite{ceperley_1980} evaluated $E_{\rm
tot,\Gamma}(N) - \Delta K(N)$, where $E_{\rm tot,\Gamma}(N)$ is the
total energy at $\Gamma$, at closed-shell system sizes in the range
$38\leq N \leq 246$, and used an extrapolation formula including a
single $N^{-1}$ term to obtain the thermodynamic limit.
A reconstruction of this approach with our DMC data is represented in
Fig.\ \ref{fig:tdl_compare} (empty diamonds and dotted line); we have
used $N^2$ as weights in the single-term fit.
Remarkably, the choice of system sizes is such that the single-term
extrapolation yields a nearly identical thermodynamic limit for
$\Gamma$-point energies as for twist averaged energies, but the
absence of higher-order terms in the extrapolation formula results in
an underestimation of the total fixed-node energy by about 2 mHa.
Higher-order contributions are less important at lower densities, and
we conclude that the extrapolation carried out by Ceperley and Alder
is very accurate at the densities for which they reported results.

The recent fixed-node DMC study of Spink \textit{et al.\@}
\cite{spink_ueg_2013} can be regarded the current state of the art in
the treatment of finite size errors.
Spink \textit{et al.\@} approximate the thermodynamic limit of the
total energy by the backflow DMC value of $E_{\rm tot}^{\rm FN}(N) -
\Delta K(N) - t_3 N^{-1} - T_4 N^{-4/3}$ at a single system size,
where the last term is the next-to-leading order contribution to the
finite size error in the DMC kinetic energy, with $T_4 =
\frac{\epsilon_3} {16\pi} r_{\rm s}^{-2}$ for the polarized electron
gas and $\epsilon_3=21.04959845$ for simple cubic simulation cells
\cite{drummond_fse_2008, supplemental}.
Our reconstruction of this approach using our (non-backflow) DMC data
is presented in Fig.\ \ref{fig:tdl_compare} (open squares).
The quasirandom fluctuations obtained with this approach are small but
still significant and, although the data extrapolate to the correct
value, individual energy values in Fig.\ \ref{fig:tdl_compare}
overestimate the thermodynamic limit by up to over $2$ mHa.
Indeed, the thermodynamic limit of the backflow DMC correlation energy
at $r_{\rm s} = 0.5$ reported by Spink \textit{et al.\@}, obtained for
a 118-electron system in a face-centred cubic simulation cell, is
$2.27(2)$ mHa above our estimate of the thermodynamic limit of the
(non-backflow) DMC correlation energy.

\begin{figure}[!hbt]
  \begin{center}
    \includegraphics[width=\columnwidth]{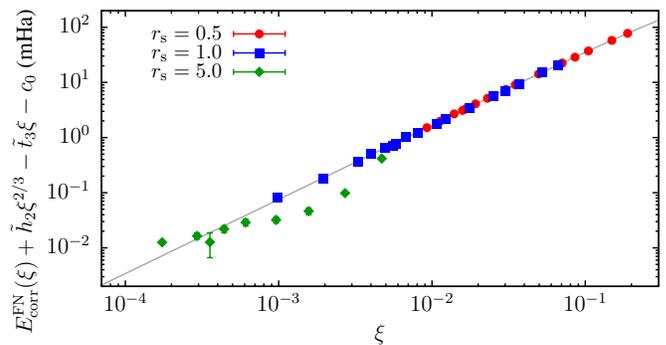}
    \caption{
      Fixed node correlation energies of the polarized UEG at $r_{\rm
      s} = 0.5$, $1$, and $5$ relative to the thermodynamic limit as a
      function of $\xi$.
      The line represents a combined fit of the data at $r_{\rm s} =
      0.5$ and $1$ to Eq.\ \ref{eq:dmc_ecorr_tdl_xi}, with
      density-dependent $c_0$ and density-independent ${\tilde c}_4$,
      ${\tilde c}_5$, and ${\tilde c}_6$ coefficients.}
    \label{fig:tdl_sjdmc_xi}
  \end{center}
\end{figure}

We turn our attention to the density dependence of Eq.\
\ref{eq:dmc_ecorr_tdl}, which we re-express as
\begin{equation}
  \label{eq:dmc_ecorr_tdl_xi}
  E_{\rm corr}^{\rm FN}(\xi)
      + {\tilde h}_2 \xi^{2/3} - {\tilde t}_3 \xi = c_0
      + {\tilde c}_4 \xi^{4/3} + {\tilde c}_5 \xi^{5/3}
      + {\tilde c}_6 \xi^2 + \ldots \;,
\end{equation}
where $\xi=r_{\rm s}^{-3/2} N^{-1}$.
We find that assuming tilded coefficients to be density-independent,
in line with leading-order extrapolation formulas proposed in the
literature \cite{ceperley_1978}, incurs a negligible error at high
densities.
In Fig.\ \ref{fig:tdl_sjdmc_xi} we plot $E_{\rm corr}^{\rm FN}(\xi)$
and perform a combined fit of the data at $r_{\rm s}=0.5$ and $1$ to
Eq.\ \ref{eq:dmc_ecorr_tdl_xi}, which we find to fit the data very
well \cite{supplemental}.
We also plot fixed node energies at $r_{\rm s}=5$ to demonstrate the
breakdown of this approximation at low densities.

We compute the exact energy of the system using FCIQMC at system sizes
$N=15$, $19$, and $27$ at $r_{\rm s}=1$ and $N=15$, $19$, $27$, and
$33$ at $r_{\rm s}=0.5$, and we evaluate the fixed node error as the
difference between the fixed node and exact correlation energies.
We find the fixed node error to increase monotonically with system
size \cite{supplemental}.

\begin{figure}[!hbt]
  \begin{center}
    \includegraphics[width=\columnwidth]{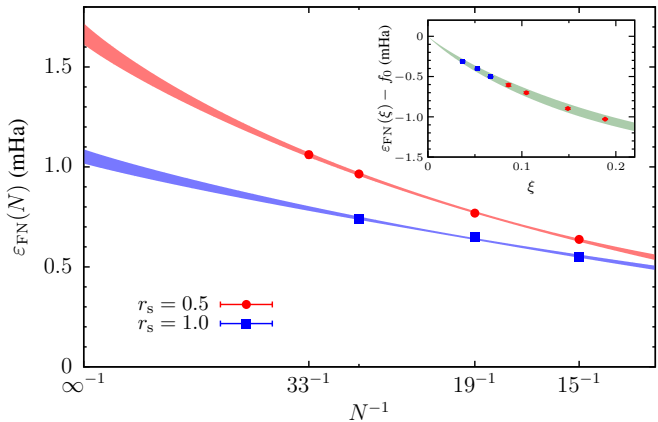}
    \caption{
      Fixed node error for the polarized UEG at $r_{\rm s} = 0.5$ and
      $1$ as a function of $N^{-1}$.
      The curves are obtained by simultaneously fitting the data at
      both densities to Eq.\ \ref{eq:fnerr_tdl} with density-dependent
      $f_0$ and density-independent ${\tilde f}_3$ and
      ${\tilde f}_4$ coefficients.
      The line width represents the statistical uncertainty in the
      fit.
      The inset shows the combined fit against $\xi$.}
    \label{fig:fnerr}
  \end{center}
\end{figure}
\begin{figure}[!hbt]
  \begin{center}
    \includegraphics[width=\columnwidth]{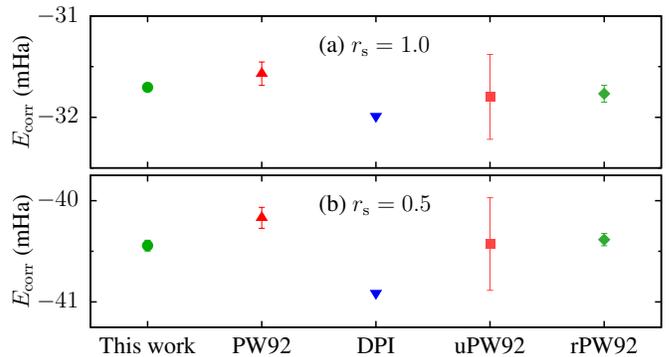}
    \caption{
      Correlation energy of the polarized UEG at (a) $r_{\rm s} = 1$
      and (b) $r_{\rm s} = 0.5$ from our caculations, and values of
      the PW92 and DPI parametrizations, our unweighted PW92 fit to
      the CA data (uPW92), and our revised unweighted PW92 fit to the
      CA data and our present results (rPW92).}
    \label{fig:final_ecorr}
  \end{center}
\end{figure}

Holzmann \textit{et al.\@} \cite{holzmann_fse_2016} found that the use
of backflow contributes to the finite size error in the energy of the
UEG at order $N^{-1}$.
This has the subtle consequence that the coefficient of $N^{-1}$ in
the finite size error of the exact energy must differ from $t_3$.
We assume the fixed node error to have the same asymptotic behavior as
the backflow contribution to the energy, which is consistent with the
observation of an approximate proportionality between these two
quantities \cite{kwon_backflow_1998, seth_atoms_2011}.
We expect $\varepsilon_{\rm FN}$ to vary less strongly with $N$ than
the fixed node energy, and thus we model it using a lower-order
expression.
Under the assumption that, like $E_{\rm corr}^{\rm FN}$, the exact
correlation energy is accurately represented at high densities by a
function of $\xi$, we write
\begin{equation}
  \label{eq:fnerr_tdl}
  \varepsilon_{\rm FN}(\xi) = f_0 + {\tilde f}_3 \xi
                 + {\tilde f}_4 \xi^{4/3} + \ldots \;,
\end{equation}
where $f_0$ is a density-dependent parameter and ${\tilde f}_3$ and
${\tilde f}_4$ are density-independent coefficients.
We perform a combined fit of our data at $r_{\rm s}=0.5$ and $1$ to
Eq.\ \ref{eq:fnerr_tdl} to obtain the thermodynamic limit of the
fixed node error at both densities.
In Fig.\ \ref{fig:fnerr} we plot the fixed node error and the
resulting fit curves, and in the inset we show the same data as a
function of $\xi$.
The results obtained with this procedure are given in Table
\ref{tab:tdl} and plotted in Fig.\ \ref{fig:final_ecorr}.

\begin{table}[!hbt]
  \begin{center}
    \begin{tabular}{lr@{.}lr@{.}l}
      \hline
        &
      \multicolumn{2}{c}{$r_{\rm s}=0.5$} &
      \multicolumn{2}{c}{$r_{\rm s}=1.0$} \\
      \hline
      \hline
      $E_{\rm corr}^{\rm FN}$
          & $-38$&$778(10)$ & $-30$&$650(3) $ \\
      $\varepsilon_{\rm FN}$
          & $  1$&$67(5)  $ & $  1$&$05(4)  $ \\
      $E_{\rm corr}$
          & $-40$&$44(5)  $ & $-31$&$70(4)  $ \\[0.2cm]
      PW92
          & $-40$&$2(1)   $ & $-31$&$6(1)   $ \\
      DPI
          & $-40$&$91     $ & $-31$&$99     $ \\
      uPW92
          & $-40$&$4(5)   $ & $-31$&$8(4)   $ \\
      rPW92
          & $-40$&$38(6)  $ & $-31$&$77(8)  $ \\
      \hline
    \end{tabular}
  \end{center}
  \caption{Thermodynamic limit of the fixed node correlation energy,
    of the fixed node error, and of the exact correlation energy of
    the polarized UEG at $r_{\rm s}=0.5$ and $1$, in mHa.
    Also shown are values of the PW92 and DPI parametrizations, an
    unweighted PW92 fit to the CA data (uPW92), and a revised
    unweighted PW92 fit to the CA data and our results (rPW92).}
  \label{tab:tdl}
\end{table}

Before comparing our results with existing parametrizations, we
note two problematic aspects of the PW92 fit.
First, the statistical uncertainty of the CA data propagates to the
parametrized correlation energies, which thus incur a random bias of
magnitude proportional to the uncertainty, but this was ignored after
fitting.
We have calculated these propagated uncertainties, shown in Table
\ref{tab:tdl} and Fig.\ \ref{fig:final_ecorr}.
Second, the CA data were weighted by their inverse square uncertainty
in the PW92 fit (a ``chi-square'' fit), but these span over two orders
of magnitude, and in effect the PW92 parametrization ignores the CA
data for $r_{\rm s}\leq 10$:\@ fitting the CA energies for $r_{\rm
s}=20$, $50$, and $100$ to the PW92 form gives essentially identical
results to the ``chi-square'' fit using all the data.
In Table \ref{tab:tdl} and Fig.\ \ref{fig:final_ecorr} we report
values of an unweighted PW92 fit to the CA data (uPW92) and of an
unweighted PW92 fit to the CA data and our present results (rPW92).

We find that the magnitude of the correlation energy is underestimated
by the PW92 parametrization by about $3$--$6$ meV, and overestimated
by the DPI parametrization by $8$--$13$ meV.
The correlation energies obtained from the unweighted uPW92 fit have
rather large uncertainties, but their expected values are more
accurate than those from the weighted fit.
Our revised rPW92 fit delivers the correct correlation energies at
both densities with negligible bias and a factor of $5$--$10$ smaller
uncertainties than the uPW92 fit.

By construction, the accuracy of the DPI parameterization at finite
densities depends exclusively on its functional form.
Modifications to include more high-density constraints would be
advisable in order to enable better agreement with our results.
Alternatively, additional degrees of freedom could be used to fit
parameters to quantum Monte Carlo data, which would be advantageous
over our rPW92 fit since the DPI form has the correct analytic
structure in the high- and low-density limits \cite{sun_dpi_2010}.


\begin{acknowledgments}
The authors would like to thank H. Luo, J.J. Shepherd, R.J. Needs,
and M. Holzmann for useful discussions.
Financial support and computational resources have been provided by
the Max-Planck-Gesellschaft.
\end{acknowledgments}


\setcounter{table}{0}
\renewcommand{\thetable}{S\arabic{table}}
\setcounter{figure}{0}
\renewcommand{\thefigure}{S\arabic{figure}}
\setcounter{section}{0}
\renewcommand{\thesection}{S\arabic{section}}
\setcounter{equation}{0}
\renewcommand{\theequation}{S\arabic{equation}}
\renewcommand{\bibnumfmt}[1]{[S#1]}
\renewcommand{\citenumfont}[1]{S#1}

\onecolumngrid
\pagebreak
\begin{center}
\textbf{\large Correlation energies of the high-density
spin-polarized electron gas to meV accuracy:\@ Supplemental Material}
\end{center}
\twocolumngrid

\section{The uniform electron gas}

The first-quantized Hamiltonian of the infinite UEG is, in Hartree
atomic units ($\hbar = m_e = |e| = 4\pi\epsilon_0 = 1$),
\begin{equation}
  \label{Seq:hamiltonian_r}
  {\hat H} = -\frac 1 2 \sum_i \nabla_i^2
           + \sum_{i < j}
               \frac 1 {\left| {\bf r}_i - {\bf r}_j\right|} \;,
\end{equation}
where ${\bf r}_i$ is the position vector of the $i$th electron, and
the system is characterized by its uniform number density $n$, usually
specified via $r_{\rm s} = \left(4\pi n/3\right)^{-1/3}$.
The second-quantized Hamiltonian of the infinite UEG is
\begin{equation}
  \label{Seq:hamiltonian_k}
  {\hat H} = \frac 1 2 \sum_{\kvec} k^2 a^\dagger_{\kvec} a_{\kvec}
           + \sum_{\pvec,\qvec} \sum_{\kvec\neq{\bm 0}}
               \frac {4\pi}{k^2} a^\dagger_{\pvec+\kvec}
               a^\dagger_{\qvec-\kvec} a_{\qvec} a_{\pvec} \;,
\end{equation}
where $\kvec$, $\pvec$, and $\qvec$ are reciprocal-space vectors, and
$a^\dagger_{\kvec}$ and $a_{\kvec}$ are the creation and annihilation
operators for the single-electron state of wave vector $\kvec$,
respectively.
The Fermi wave vector, $k_{\rm F} = (6\pi^2 n)^{1/3}$ at full spin
polarization, characterizes the system.
The kinetic energy term is diagonal, and the interaction term only
connects states with equal total momentum $\kvec_{\rm T}$.
The Hilbert space of the system thus consists of disjoint subspaces
corresponding to different $\kvec_{\rm T}$, and the ground state is
the solution of the Schr\"odinger equation in the subspace for which
the total energy is minimized.

We simulate a finite version of this system consisting of $N$
electrons in a cubic simulation cell of side $L=(n/N)^{1/3}$ subject
to periodic boundary conditions.
This requires replacing the Coulomb interaction in Eq.\
\ref{Seq:hamiltonian_r} with an Ewald summation
\cite{suppl_ewald_1921}, restricting the summations in Eq.\
\ref{Seq:hamiltonian_k} to reciprocal lattice vectors, $\Gvec =
\frac{2\pi} L (i_x,i_y,i_z)$, where $i_x$, $i_y$, and $i_z$ are
integers, and adding a self-interaction constant to both Hamiltonians.

In the high density regime the UEG behaves as a Fermi liquid, for
which a plane-wave basis is a natural choice.
The configuration interaction (CI) expansion of the ground-state wave
function is $\Psi_0 = \sum_I C_I D_I$, where $\{C_I\}$ are the CI
coefficients, $D_I = \det(e^{i \Gvec_{\mu_{Ij}}\cdot\rvec_i})$ are
determinants of plane-wave orbitals, and $\mu_{Ij}$ is the index of
the $j$th wave vector occupied in the $I$th determinant.
We label the HF determinant, which corresponds to the choice of $I$
that minimizes $\langle D_I\vert {\hat H} \vert D_I\rangle / \langle
D_I\vert D_I\rangle$, as $I=1$.

\section{Twist averaging}

The translational invariance of the wave function of a periodic system
is defined up to a phase factor, $\Psi(\rvec_1, ..., \rvec_i+\Rvec,
...,\rvec_N) = e^{i\theta}\Psi(\rvec_1, ..., \rvec_i, ...,\rvec_N)$,
where $\Rvec$ is a simulation cell lattice vector, $\Rvec =
L(i_x,i_y,i_z)$, and $i_x$, $i_y$, and $i_z$ are integers.
This phase factor can be obtained by shifting the reciprocal lattice
by a certain $\kvec_{\rm s}$ in the Brillouin zone such that $\theta =
\kvec_{\rm s}\cdot\Rvec$.

We note that the total momentum $\kvec_{\rm T} = \sum_i
\Gvec_{\mu_{1i}}$ of the ground-state wave function changes discretely
with $\kvec_{\rm s}$, dividing the Brillouin zone into $Z$ regions
associated with different $\kvec_{\rm T}$.
Since it is not trivial to determine \textit{a priori} which
$\kvec_{\rm T}$ yields the lowest energy at a given $\kvec_{\rm s}$,
$\kvec_{\rm T}$ is usually chosen so as to minimize the energy of the
non-interacting system, resulting in convex polyhedral regions
bounded by Bragg planes \cite{suppl_lin_ta_2001}.

Averaging an expectation value $A$ over $\kvec_{\rm s}$ in the
Brillouin zone,
\begin{equation}
  \label{Seq:twist-averaging}
  A_{\rm TA} =
    \frac 1 \Omega_{\rm BZ}
    \int_{\rm BZ} A(\kvec_{\rm s})
         \,{\rm d}\kvec_{\rm s} \;,
\end{equation}
where $\Omega_{\rm BZ}$ is the volume of the Brillouin zone, is
referred to as twist averaging, and has the effect of reducing
quasirandom fluctuations of the expectation value with system size
$N$ \cite{suppl_lin_ta_2001}.
The integration over $\kvec_{\rm s}$ is usually performed
stochastically or using a grid in the Brillouin zone
\cite{suppl_lin_ta_2001, suppl_drummond_fse_2008}.
However, inspection of the second-quantized Hamiltonian of Eq.\
\ref{Seq:hamiltonian_k} reveals that, for a fixed $\kvec_{\rm T}$,
shifting the reciprocal lattice by $\kvec_{\rm s}$ adds a constant to
the diagonal kinetic energy term and leaves the interaction term
unchanged, since it only depends on differences between reciprocal
lattice vectors.
Therefore the correlation energy only depends on the total momentum
$\kvec_{\rm T}$, which changes discretely with $\kvec_{\rm s}$, and
therefore the integral reduces to a sum over the $Z$ regions in which
the total momentum is constant,
\begin{equation}
  \label{Seq:tabc}
  \begin{split}
  E_{\rm corr}^{\rm TA} & =
    \frac 1 \Omega_{\rm BZ}
    \int_{\rm BZ} E_{\rm corr}(\kvec_{\rm s})
         \,{\rm d}\kvec_{\rm s} \\
    & = \sum_z \frac{\Omega_z}{\Omega_{\rm BZ}}
               E_{\rm corr}(\kvec_{\rm s}^z) \;,
  \end{split}
\end{equation}
where $\Omega_z$ is the volume of the $z$th region and $\kvec_{\rm
s}^z$ is an arbitrary reciprocal lattice shift in the $z$th region.

This suggests a twist-averaging scheme which is more efficient than
other approaches at small system sizes
\cite{suppl_shepherd_thesis_2013}.
By expressing the total energy as the HF energy plus the correlation
energy, $E_{\rm tot}(\kvec_{\rm s}) = E_{\rm HF}(\kvec_{\rm s}) +
E_{\rm corr}(\kvec_{\rm s})$, the HF energy absorbs the continuous
variation of the kinetic energy with $\kvec_{\rm s}$, while the
correlation energy is constant within each region.
Evaluating the average correlation energy weighted by the region
volumes, which can be obtained exactly for $N \lesssim 100$, see
below, yields the twist-averaged correlation energy.
We use this scheme to twist-average our FCIQMC energies, and we use
random sampling to twist-average our DMC energies.

\subsection{Exact division of the Brillouin zone}

The energy (per electron) of the non-interacting electron gas equals
the HF kinetic energy,
\begin{equation}
  \label{Seq:E_NI}
  \begin{split}
  E_{\rm NI}(\kvec_{\rm s};\kvec_{\rm T}) &
      = K_1(\kvec_{\rm s};\kvec_{\rm T}) \\
    & = \frac 1 {2N} \sum_{i=1}^N (\Gvec_{\mu_{1i}}+\kvec_{\rm s})^2 \\
    & = \frac 1 {2N} \sum_{i=1}^N (G_{\mu_{1i}}^2+k_s^2+2
                         \Gvec_{\mu_{1i}}\cdot\kvec_{\rm s}) \\
    & = E_{\rm NI}({\bf 0};\kvec_{\rm T}) + \frac 1 2 k_s^2
      + \frac 1 N \kvec_{\rm T}\cdot\kvec_{\rm s}
    \;,
  \end{split}
\end{equation}
where $\{\mu_{1i}\}$ are the indices of the reciprocal lattice vectors
occupied in the HF determinant.
These indices determine $\kvec_{\rm T}$ and \textit{vice versa}.
The energy of the non-interacting system at fixed $\kvec_{\rm T}$ is
a paraboloid centred at $\kvec_{\rm s}=-\frac 1 N \kvec_{\rm T}$.
Since the total momentum at shift $\kvec_{\rm s}$ is that which
minimizes $E_{\rm NI}(\kvec_{\rm s};\kvec_{\rm T})$, $\kvec_{\rm T}$
changes discretely at the intersection of two such paraboloids.
If $\kvec_{\rm T}^{z_1}$ and $\kvec_{\rm T}^{z_2}$ are the total
momenta of two adjacent regions, this intersection is given by
\begin{equation}
  \label{Seq:E_NI_intersection}
  \frac 1 N \left( \kvec_{\rm T}^{z_1}
                 - \kvec_{\rm T}^{z_2} \right)
            \cdot \kvec_{\rm s}
  = E_{\rm NI}({\bf 0};\kvec_{\rm T}^{z_2})
  - E_{\rm NI}({\bf 0};\kvec_{\rm T}^{z_1}) \;,
\end{equation}
which is the equation of a plane.
The Brillouin zone regions of constant total momentum are therefore
convex polyhedra.

In practice we work in the irreducible Brullouin zone (IBZ), which for
a simple cubic simulation cell is the tetrahedron given by $0\leq
z\leq y\leq x\leq \pi/L$, where $x$, $y$, and $z$ are the Cartesian
components of $\kvec_{\rm s}$.
Consequently the total momentum $\kvec_{\rm T}=\frac \pi L (i_x, i_y,
i_z)$, where $i_x$, $i_y$, and $i_z$ are integers, satisfies $0\leq
-i_z \leq -i_y \leq -i_x \leq N/2$.

The problem of dividing the IBZ reduces to locating the vertices of
the polyhedral regions.
Note that Eq.\ \ref{Seq:E_NI_intersection} represents a Bragg plane,
which can be defined in terms of integers, and the region vertices are
the intersections of three or more inter-region and/or IBZ planes, and
are therefore proportional to vectors of rational numbers.
The use of integer arithmetic enables solving the IBZ division problem
exactly for moderate system sizes.

Given a shift $\kvec_{\rm s}$, finding the $N$ reciprocal lattice
vectors with the smallest $\vert \Gvec_j+\kvec_{\rm s}\vert$ yields
the indices of the occupied orbitals $\{\mu_{1i}\}$, which determines
$\kvec_{\rm T}$.
However, at points on inter-region planes the set of occupied orbitals
is not unique, and multiple total momenta give the same, degenerate
kinetic energy.
The allowed values of the total momentum at a vertex can be obtained
by considering all possible occupations, and the equations of the
inter-region planes passing through the vertex are given by Eq.\
\ref{Seq:E_NI_intersection} for each pair of allowed total momenta.
In turn, each pair of planes intersect at a line corresponding to
a polyhedral edge which points to an adjacent vertex.

\begin{figure}[!hbt]
  \begin{center}
    \includegraphics[width=0.9\columnwidth]{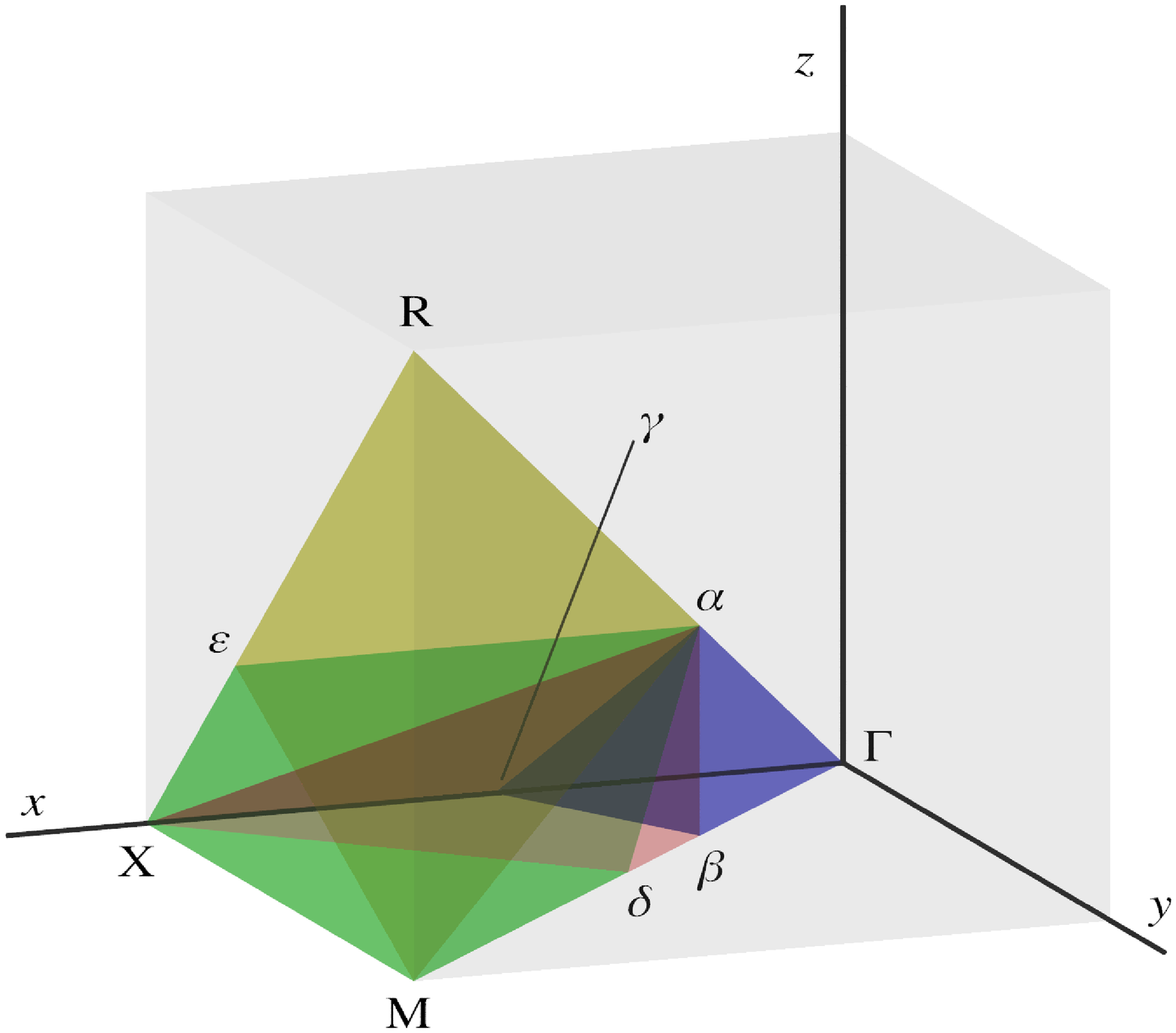}
    \caption{Division of the IBZ of the $7$-electron gas into regions
      of constant total momentum.}
    \label{Sfig:bz7}
  \end{center}
\end{figure}

It is thus possible to find the vertices of all polyhedral regions in
the IBZ by successively moving between adjacent vertices along region
edges.
We illustrate our algorithm using the particularly simple case of the
$7$-electron gas, which we do not consider in our main results.
The IBZ division for this example is shown in Fig.\ \ref{Sfig:bz7},
where we have labelled the high-symmetry points $\Gamma$, X, M, and R
at the corners of the IBZ and the additional vertices $\alpha$,
$\beta$, $\gamma$, $\delta$, and $\varepsilon$.

We start at $\Gamma$, where we find that $\kvec_{\rm T}={\bf 0}$ is
the only allowed value of the total momentum.
We then perform a line search between $\Gamma$ and M, corresponding to
the intersection between two of the three IBZ planes passing through
$\Gamma$, to find the point furthest from $\Gamma$ at which
\textit{any} of the allowed values of the total momentum at $\Gamma$
is also an allowed value.
This is done by bisection using floating-point arithmetic, and upon
locating vertex $\beta=\frac{2\pi}L (\frac 1 6,\frac 1 6,0)$ we revert
to using integer arithmetic.
Inspecting the degenerate occupations at $\beta$ yields two possible
total momenta, $\kvec_{\rm T}={\bf 0}$ and $-\frac{2\pi}L (2,1,0)$,
defining an inter-region plane of normal $(2,1,0)$.
The pairwise intersections between this plane and the two IBZ planes
passing through $\beta$ provide search directions to find adjacent
vertices $\alpha$, $\gamma$, and $\delta$, and this process continues
until we exhaust the lists of edges radiating from all vertices.

The resulting vertex locations characterize the regions and, using the
divergence theorem, we obtain their volume, center, and contributions
to the HF kinetic energy (the HF exchange energy is constant within
each region).
For instance, the twist-averaged kinetic and exchange energy of the
$7$-electron gas are exactly
\begin{equation}
  K = \frac{215}{504}
      \left(\frac{6 \pi^2}{7}\right)^{2/3}
      r_{\rm s}^{-2} \;,
\end{equation}
and
\begin{equation}
  X = -\frac{7459}{3780}
      \left(\frac{3} {28 \pi^4}\right)^{1/3}
      r_{\rm s}^{-1} + v_{\rm M} r_{\rm s}^{-1}\;,
\end{equation}
where $v_{\rm M} = -0.46005809$ is the self-image energy
contribution for this system at $r_{\rm s}=1$.
These integrals can also be carried out accurately using
floating-point arithmetic; the use of integers is however crucial for
the location of vertices, since for $N \gtrsim 30$ the proximity of
some of the vertices can cause incorrect IBZ division under
floating-point arithmetic.

\begin{figure}[!hbt]
  \begin{center}
    \includegraphics[width=\columnwidth]{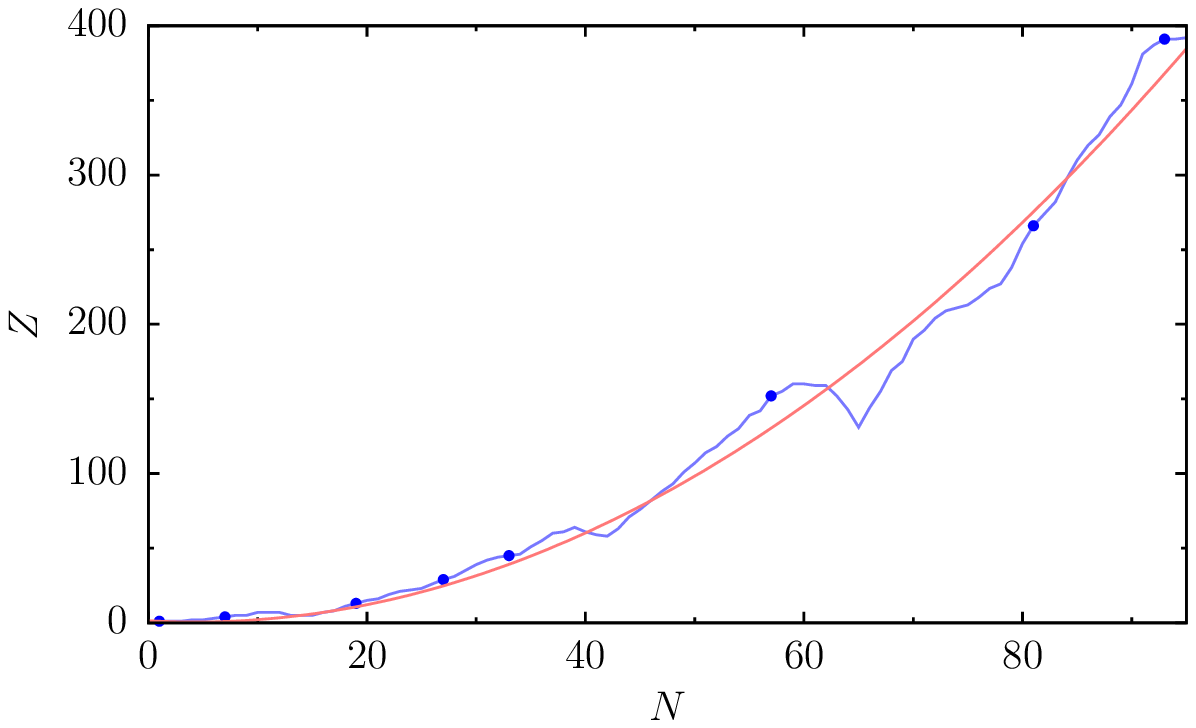}
    \caption{Number of IBZ regions as a function of system size for
      simple cubic simulation cells.
      The circles correspond to closed shell systems, and the red line
      is a fitted parabola to guide the eye.}
    \label{Sfig:nreg}
  \end{center}
\end{figure}

In Tables \ref{Stab:regions_7_15_19}, \ref{Stab:regions_27}, and
\ref{Stab:regions_33} we give the exact region volumes and centers
(the latter truncated to four decimals for conciseness) of the IBZ
regions corresponding to $N=15$, $19$, $27$, and $33$, for which we
have run FCIQMC calculations, as well as those for the $7$-electron
system of Fig.\ \ref{Sfig:bz7}.
In Fig.\ \ref{Sfig:nreg} we plot the number of regions $Z$ as a
function of $N$, showing that $Z \sim N^2$.
Therefore the number of evaluations of an expectation value required
for twist averaging increases quadratically with system size.

\begin{table}[!hbt]
  \begin{center}
    \begin{tabular}{rr@{\quad}ccc}
      \hline
      \multicolumn{1}{c}{$N$}                                &
      \multicolumn{1}{c}{$z$}                                &
      \multicolumn{1}{c}{$-\frac L{2\pi} {\bf k}_{\rm T}^z$} &
      \multicolumn{1}{c}{$\Omega_z/\Omega_{\rm BZ}$}         &
      \multicolumn{1}{c}{$\frac L{2\pi} {\bf k}_{\rm s}^z$}  \\
      \hline
      \hline
      $ 7$ & $ 0$ & $(0,0,0)$ & $    1/18$
                              & $(0.1458,0.0833,0.0417)$ \\
           & $ 1$ & $(2,1,0)$ & $     1/9$
                              & $(0.2708,0.1146,0.0417)$ \\
           & $ 2$ & $(3,2,1)$ & $    7/18$
                              & $(0.3899,0.2173,0.0655)$ \\
           & $ 3$ & $(3,3,3)$ & $     4/9$
                              & $(0.4167,0.3333,0.2083)$ \\[0.1cm]
      $15$ & $ 0$ & $(3,2,1)$ & $  13/252$
                              & $(0.1513,0.1036,0.0499)$ \\
           & $ 1$ & $(4,0,0)$ & $    1/28$
                              & $(0.2411,0.0491,0.0179)$ \\
           & $ 2$ & $(5,2,2)$ & $    4/21$
                              & $(0.3452,0.1845,0.1533)$ \\
           & $ 3$ & $(4,4,1)$ & $    7/72$
                              & $(0.2798,0.2485,0.1488)$ \\
           & $ 4$ & $(6,4,0)$ & $     5/8$
                              & $(0.4250,0.2937,0.1250)$ \\[0.1cm]
      $19$ & $ 0$ & $(0,0,0)$ & $    1/40$
                              & $(0.1188,0.0562,0.0250)$ \\
           & $ 1$ & $(2,2,1)$ & $  11/360$
                              & $(0.1680,0.1055,0.0553)$ \\
           & $ 2$ & $(4,3,1)$ & $     1/7$
                              & $(0.2606,0.1728,0.0558)$ \\
           & $ 3$ & $(6,2,1)$ & $ 83/2520$
                              & $(0.3354,0.1143,0.0497)$ \\
           & $ 4$ & $(5,3,3)$ & $  19/105$
                              & $(0.3392,0.2646,0.2079)$ \\
           & $ 5$ & $(6,4,2)$ & $    4/55$
                              & $(0.3856,0.3391,0.1614)$ \\
           & $ 6$ & $(7,3,0)$ & $205/5544$
                              & $(0.3920,0.2406,0.0216)$ \\
           & $ 7$ & $(8,0,0)$ & $    1/56$
                              & $(0.3705,0.0491,0.0179)$ \\
           & $ 8$ & $(8,3,2)$ & $ 146/693$
                              & $(0.4487,0.2780,0.1765)$ \\
           & $ 9$ & $(9,1,1)$ & $  43/504$
                              & $(0.4479,0.1271,0.0540)$ \\
           & $10$ & $(7,6,1)$ & $  19/792$
                              & $(0.4098,0.3794,0.1298)$ \\
           & $11$ & $(9,5,1)$ & $  17/396$
                              & $(0.4711,0.3528,0.1290)$ \\
           & $12$ & $(9,8,0)$ & $    7/72$
                              & $(0.4658,0.4211,0.1250)$ \\[0.1cm]
      \hline
    \end{tabular}
  \end{center}
  \caption{Index $z$, total momentum ${\bf k}_{\rm T}^z$, exact weight
    $\Omega_z/\Omega_{\rm BZ}$, and center ${\bf k}_{\rm s}^z$ of the
    IBZ regions for $N=7$, $15$, and $19$.}
  \label{Stab:regions_7_15_19}
\end{table}

\begin{table}[!hbt]
  \begin{center}
    \begin{tabular}{rr@{\quad}ccc}
      \hline
      \multicolumn{1}{c}{$N$}                                &
      \multicolumn{1}{c}{$z$}                                &
      \multicolumn{1}{c}{$-\frac L{2\pi} {\bf k}_{\rm T}^z$} &
      \multicolumn{1}{c}{$\Omega_z/\Omega_{\rm BZ}$}         &
      \multicolumn{1}{c}{$\frac L{2\pi} {\bf k}_{\rm s}^z$}  \\
      \hline
      \hline
      $27$ & $ 0$ & $( 0, 0, 0)$ & $       1/60$ &
                                 $(0.0979,0.0563,0.0250)$ \\
           & $ 1$ & $( 3, 1, 1)$ & $       2/35$ &
                                 $(0.2009,0.0828,0.0493)$ \\
           & $ 2$ & $( 4, 4, 0)$ & $ 2363/18900$ &
                                 $(0.2514,0.1808,0.0633)$ \\
           & $ 3$ & $( 5, 3, 3)$ & $     26/945$ &
                                 $(0.2706,0.2041,0.1763)$ \\
           & $ 4$ & $( 6, 3, 0)$ & $       1/84$ &
                                 $(0.3235,0.0908,0.0250)$ \\
           & $ 5$ & $( 7, 4, 1)$ & $ 1409/17160$ &
                                 $(0.3423,0.2085,0.0910)$ \\
           & $ 6$ & $( 6, 5, 4)$ & $    34/2925$ &
                                 $(0.3055,0.2652,0.2258)$ \\
           & $ 7$ & $( 7, 6, 0)$ & $      1/200$ &
                                 $(0.3271,0.2889,0.0250)$ \\
           & $ 8$ & $( 9, 2, 2)$ & $     16/315$ &
                                 $(0.3851,0.1327,0.0968)$ \\
           & $ 9$ & $( 7, 7, 2)$ & $      2/585$ &
                                 $(0.3143,0.2951,0.1647)$ \\
           & $10$ & $( 9, 5, 0)$ & $      1/840$ &
                                 $(0.4051,0.2286,0.0114)$ \\
           & $11$ & $(10, 3, 1)$ & $  299/18480$ &
                                 $(0.4258,0.1805,0.0586)$ \\
           & $12$ & $(10, 5, 2)$ & $ 1732/69615$ &
                                 $(0.4111,0.2256,0.1476)$ \\
           & $13$ & $( 9, 7, 1)$ & $ 1009/10296$ &
                                 $(0.3923,0.3282,0.0672)$ \\
           & $14$ & $( 9, 6, 5)$ & $    33/7735$ &
                                 $(0.3493,0.2623,0.2338)$ \\
           & $15$ & $( 8, 8, 4)$ & $  298/47775$ &
                                 $(0.3307,0.3078,0.2195)$ \\
           & $16$ & $(12, 1, 0)$ & $       1/16$ &
                                 $(0.4485,0.1013,0.0417)$ \\
           & $17$ & $(11, 5, 5)$ & $    11/1428$ &
                                 $(0.4264,0.2296,0.2149)$ \\
           & $18$ & $(10, 8, 3)$ & $ 2458/97461$ &
                                 $(0.3989,0.3563,0.1501)$ \\
           & $19$ & $(12, 6, 1)$ & $    62/1001$ &
                                 $(0.4670,0.2841,0.0559)$ \\
           & $20$ & $(13, 4, 0)$ & $      5/504$ &
                                 $(0.4790,0.1963,0.0417)$ \\
           & $21$ & $(12, 7, 3)$ & $   145/1989$ &
                                 $(0.4609,0.3058,0.1475)$ \\
           & $22$ & $(11, 8, 6)$ & $8413/324870$ &
                                 $(0.3955,0.2993,0.2390)$ \\
           & $23$ & $(11,10, 5)$ & $    23/3185$ &
                                 $(0.4239,0.3977,0.1563)$ \\
           & $24$ & $(13, 9, 5)$ & $     10/637$ &
                                 $(0.4754,0.3651,0.1677)$ \\
           & $25$ & $(12, 9, 9)$ & $       1/30$ &
                                 $(0.4292,0.3458,0.3250)$ \\
           & $26$ & $(11,11, 8)$ & $      5/147$ &
                                 $(0.4066,0.3829,0.3036)$ \\
           & $27$ & $(13,10, 8)$ & $     19/588$ &
                                 $(0.4745,0.3639,0.2979)$ \\
           & $28$ & $(13,12, 7)$ & $     17/245$ &
                                 $(0.4718,0.4330,0.2307)$ \\
      \hline
    \end{tabular}
  \end{center}
  \caption{Index $z$, total momentum ${\bf k}_{\rm T}^z$, exact weight
    $\Omega_z/\Omega_{\rm BZ}$, and center ${\bf k}_{\rm s}^z$ of the
    IBZ regions for $N=27$.}
  \label{Stab:regions_27}
\end{table}

\begin{table}[!hbt]
  \begin{center}
    \begin{tabular}{rr@{\quad}ccc}
      \hline
      \multicolumn{1}{c}{$N$}                                &
      \multicolumn{1}{c}{$z$}                                &
      \multicolumn{1}{c}{$-\frac L{2\pi} {\bf k}_{\rm T}^z$} &
      \multicolumn{1}{c}{$\Omega_z/\Omega_{\rm BZ}$}         &
      \multicolumn{1}{c}{$\frac L{2\pi} {\bf k}_{\rm s}^z$}  \\
      \hline
      \hline
      $33$ & $ 0$ & $( 0, 0, 0)$ & $           1/100$
                                 & $(0.0813,0.0500,0.0250)$ \\
           & $ 1$ & $( 4, 1, 0)$ & $            1/75$
                                 & $(0.1464,0.0552,0.0250)$ \\
           & $ 2$ & $( 6, 3, 1)$ & $          13/840$
                                 & $(0.1972,0.0757,0.0332)$ \\
           & $ 3$ & $( 5, 5, 0)$ & $          11/600$
                                 & $(0.1712,0.1362,0.0250)$ \\
           & $ 4$ & $( 8, 3, 2)$ & $   25343/1345960$
                                 & $(0.2743,0.0781,0.0359)$ \\
           & $ 5$ & $( 7, 5, 3)$ & $      1787/29260$
                                 & $(0.2193,0.1475,0.0730)$ \\
           & $ 6$ & $(10, 2, 0)$ & $       301/22770$
                                 & $(0.3483,0.0460,0.0122)$ \\
           & $ 7$ & $( 9, 5, 0)$ & $     1889/471960$
                                 & $(0.2917,0.1500,0.0073)$ \\
           & $ 8$ & $(11, 3, 3)$ & $       549/32890$
                                 & $(0.3710,0.0763,0.0526)$ \\
           & $ 9$ & $(10, 6, 3)$ & $3734168/98423325$
                                 & $(0.3157,0.1729,0.0591)$ \\
           & $10$ & $( 8, 8, 5)$ & $            2/55$
                                 & $(0.2612,0.2293,0.1195)$ \\
           & $11$ & $( 9, 6, 6)$ & $      1007/24255$
                                 & $(0.2953,0.1833,0.1524)$ \\
           & $12$ & $(12, 5, 1)$ & $         31/8073$
                                 & $(0.3750,0.0972,0.0110)$ \\
           & $13$ & $(11, 7, 0)$ & $           1/600$
                                 & $(0.3246,0.2217,0.0050)$ \\
           & $14$ & $(13, 6, 2)$ & $    27817/464100$
                                 & $(0.4036,0.1435,0.0467)$ \\
           & $15$ & $(12, 7, 4)$ & $       347/36036$
                                 & $(0.3778,0.1955,0.1213)$ \\
           & $16$ & $(11, 9, 3)$ & $         73/2100$
                                 & $(0.3364,0.2701,0.0601)$ \\
           & $17$ & $(10, 9, 6)$ & $         65/1764$
                                 & $(0.3280,0.2704,0.1768)$ \\
           & $18$ & $(11,11, 0)$ & $         37/2520$
                                 & $(0.3578,0.3237,0.0187)$ \\
           & $19$ & $(13, 8, 3)$ & $      2423/97240$
                                 & $(0.4171,0.2540,0.0893)$ \\
           & $20$ & $(12, 8, 6)$ & $       199/24255$
                                 & $(0.3875,0.2281,0.1646)$ \\
           & $21$ & $(11,10, 5)$ & $    14593/556920$
                                 & $(0.3754,0.3202,0.1341)$ \\
           & $22$ & $(13,10, 2)$ & $     6613/245700$
                                 & $(0.4160,0.3218,0.0587)$ \\
           & $23$ & $(15, 7, 1)$ & $          29/680$
                                 & $(0.4627,0.1954,0.0341)$ \\
           & $24$ & $(15, 5, 5)$ & $          11/702$
                                 & $(0.4445,0.1286,0.1093)$ \\
           & $25$ & $(11, 9, 9)$ & $         34/2205$
                                 & $(0.3324,0.2836,0.2479)$ \\
           & $26$ & $(14, 9, 5)$ & $         9/11900$
                                 & $(0.4532,0.3040,0.1315)$ \\
           & $27$ & $(12,12, 4)$ & $         17/8190$
                                 & $(0.3904,0.3712,0.0948)$ \\
           & $28$ & $(15, 9, 0)$ & $          13/840$
                                 & $(0.4605,0.2957,0.0250)$ \\
           & $29$ & $(16, 6, 4)$ & $          16/663$
                                 & $(0.4786,0.1662,0.0926)$ \\
           & $30$ & $(15, 7, 6)$ & $   97021/2702700$
                                 & $(0.4584,0.2306,0.1637)$ \\
           & $31$ & $(13, 9, 8)$ & $      1027/29988$
                                 & $(0.4118,0.2758,0.2103)$ \\
           & $32$ & $(14,11, 4)$ & $          7/8398$
                                 & $(0.4516,0.3239,0.1226)$ \\
           & $33$ & $(14,13, 1)$ & $           1/189$
                                 & $(0.4327,0.4082,0.0172)$ \\
           & $34$ & $(13,13, 6)$ & $    15361/881790$
                                 & $(0.4234,0.3905,0.1528)$ \\
           & $35$ & $(12,11,11)$ & $           2/315$
                                 & $(0.3532,0.3171,0.2813)$ \\
           & $36$ & $(16,12, 1)$ & $            1/75$
                                 & $(0.4774,0.3750,0.0274)$ \\
           & $37$ & $(15,13, 3)$ & $     8188/166725$
                                 & $(0.4644,0.4082,0.0853)$ \\
           & $38$ & $(16,10, 7)$ & $         13/2550$
                                 & $(0.4824,0.2998,0.1650)$ \\
           & $39$ & $(14,11,10)$ & $           1/180$
                                 & $(0.4145,0.3022,0.2575)$ \\
           & $40$ & $(16,12, 6)$ & $         20/2907$
                                 & $(0.4821,0.3647,0.1430)$ \\
           & $41$ & $(15,12, 9)$ & $         79/2142$
                                 & $(0.4546,0.3571,0.1979)$ \\
           & $42$ & $(14,14,12)$ & $            1/63$
                                 & $(0.4018,0.3726,0.2494)$ \\
           & $43$ & $(16,13,12)$ & $          17/900$
                                 & $(0.4689,0.3400,0.2505)$ \\
           & $44$ & $(16,15,15)$ & $          22/225$
                                 & $(0.4561,0.4038,0.3205)$ \\
      \hline
    \end{tabular}
  \end{center}
  \caption{Index $z$, total momentum ${\bf k}_{\rm T}^z$, exact weight
    $\Omega_z/\Omega_{\rm BZ}$, and center ${\bf k}_{\rm s}^z$ of the
    IBZ regions for $N=33$.}
  \label{Stab:regions_33}
\end{table}

The need to enumerate all possible occupations of partially-filled
shells causes a computational bottleneck in our exact division
algorithm, which we are able to use in practice for $N\lesssim 100$.
Computing twist-averaged DMC correlation energies requires knowledge
of the corresponding twist-averaged HF energy components, which we
obtain using random sampling for $N \gtrsim 100$.
Our twist-averaged HF energies are given in Table \ref{Stab:tahf}.

\begin{table}[!hbt]
  \begin{center}
    \begin{tabular}{rll}
      \hline
      \multicolumn{1}{c}{$N$}             &
      \multicolumn{1}{c}{$r_{\rm s}^2 K$} &
      \multicolumn{1}{c}{$r_{\rm s} X$}   \\
      \hline
      \hline
      $     7$ & $1.77110059  $ & $-0.663751377  $ \\
      $    15$ & $1.75971498  $ & $-0.630999714  $ \\
      $    19$ & $1.75843687  $ & $-0.623184756  $ \\
      $    27$ & $1.75774258  $ & $-0.613700247  $ \\
      $    33$ & $1.75826227  $ & $-0.608535468  $ \\
      $    40$ & $1.75615221  $ & $-0.605364222  $ \\
      $    57$ & $1.75545710  $ & $-0.599501435  $ \\
      $    81$ & $1.75453662  $ & $-0.595096762  $ \\
      $    93$ & $1.75476609  $ & $-0.593397250  $ \\
      $   123$ & $1.7545311(2)$ & $-0.59069020(3)$ \\
      $   147$ & $1.7542302(2)$ & $-0.58928659(4)$ \\
      $   171$ & $1.7544501(1)$ & $-0.58801682(4)$ \\
      $   179$ & $1.7544764(2)$ & $-0.58767775(4)$ \\
      $   203$ & $1.7541746(2)$ & $-0.58696564(3)$ \\
      $   251$ & $1.7542188(1)$ & $-0.58565017(4)$ \\
      $   305$ & $1.7541073(2)$ & $-0.58465873(4)$ \\
      $   515$ & $1.7541090(1)$ & $-0.58245938(5)$ \\
      $  1021$ & $1.7540160(2)$ & $-0.58058038(5)$ \\
      $  2007$ & $1.7540110(2)$ & $-0.57937257(5)$ \\
      $\infty$ & $1.75399969  $ & $-0.577252097  $ \\
      \hline
    \end{tabular}
  \end{center}
  \caption{Twist-averaged HF kinetic and exchange energies for the
    polarized UEG at several system sizes, in Ha.
    Energies for $N\leq 93$ are exact, and energies for $N\geq 123$
    have been estimated using random sampling.
    The analytic $N\to\infty$ limit is also shown, for reference.}
  \label{Stab:tahf}
\end{table}

\section{Fitting methodology}

In our work we use least-squares fits of energy data to perform
extrapolations with respect to basis-set size and system size, as well
as to handle parametrizations of the correlation energy.
We avoid the use of ``chi-square'' fits in which each datum is
weighed by the inverse of its squared uncertainty, since this
distorts the relative importance of the data, in turn causing an
underestimation of the uncertainty in functions of the fit.
Instead we perform our least squares fits without these weights, and
we obtain uncertainties by a stochastic process in which we replace
each datum with a random number drawn from a normal distribution
centred at its expected value of variance its standard error.
The standard error in a function of the fit is then obtained as the
square root of the variance of the values of the function in 10,000
realizations of this process.

We note that the statistical uncertainty in functions of the fit
merely reflects that in the input data.
The statistical uncertainty does not capture the bias due to the
choice of a specific fitting function, which we refer to as
\textit{parametrization bias}, and has little practical meaning in the
presence of large quasirandom fluctuations.
We treat our energy data so that quasirandom fluctuations are almost
negligible, and we use fitting functions with more parameters than
strictly required for an accurate fit in order to account for part of
the parametrization bias.
While this is not a rigorous approach, we expect our estimated
statistical uncertainties to be at worst of the same order of
magnitude as the true uncertainties.

\section{Variational and diffusion Monte Carlo}

The VMC method \cite{suppl_mcmillan_1965, suppl_foulkes_rmp_2001}
requires a trial wave function $\Psi_{\rm T}$ to evaluate $\langle
\Psi_{\rm T} \vert {\hat H} \vert \Psi_{\rm T} \rangle / \langle
\Psi_{\rm T} \vert \Psi_{\rm T} \rangle$ by direct Monte Carlo
integration in real space, and provides a framework for optimizing
wave function parameters \cite{suppl_toulouse_emin_2007,
suppl_umrigar_emin_2007}.
In the DMC method \cite{suppl_anderson_1976, suppl_foulkes_rmp_2001}
the wave function is represented by a set of real-space walkers which
evolves according to a small time-step approximation
\cite{suppl_zen_2016} to the Green's function associated with the
imaginary-time Schr\"odinger equation.
The fixed node approximation prevents this process from collapsing
onto the bosonic ground state by requiring the DMC wave function to
have the same nodes as $\Psi_{\rm T}$.
The positive bias in the energy incurred by the fixed node
approximation is referred to as the fixed node error,
$\varepsilon_{\rm FN}$.

All of our VMC and DMC calculations have been performed using the
\textsc{casino} code \cite{suppl_casino}.
Each of our DMC energies is obtained by linear extrapolation of the
results of a DMC calculation consisting of $M_1$ steps with a time
step of $\tau_1$ and a target walker population of $P_1$, and a second
DMC calculation consisting of $M_2=M_1/2$ steps with a time step of
$\tau_2=4\tau_1$ and a target walker population of $P_2=P_1/4$.
We set $\tau_1 = 0.01 r_{\rm s}^2$, $P_1=2048$ walkers, and adjust
$M_1$ to obtain the desired statistical accuracy.

For our twist averaged VMC calculations we have used 6400 random
values of ${\bf k}_{\rm s}$, and for our twist averaged DMC
calculations we have used up to 3200 values for the system sizes at
which we compute the fixed node error, and 32 values for other system
sizes.

\subsection{Trial wave functions}

The Slater-Jastrow form is a common choice of trial wave function for
electronic systems, and consists of the HF determinant multiplied by a
Jastrow correlation factor, $\Psi_{\rm T}({\bf R}) = e^{J({\bf R})}
\Psi_{\rm HF}({\bf R})$.
We parametrize $J({\bf R})$ as
\cite{suppl_drummond_jastrow_2004, suppl_lopezrios_jastrow_2012}
\begin{equation}
  \label{Seq:jastrow_dtn}
  \begin{split}
    J({\bf R}) & =
    \sum_{i<j} \left(1-r_{ij}/L_u\right)^3
    \Theta\left(r_{ij}-L_u\right)
    \sum_{l=0}^8 \alpha_l r_{ij}^l \\
    & +
    \sum_{i<j}
    \sum_{s=1}^8 a_s \sum_{{\bf G} \in s{\rm th~star}}
    \cos\left({\bf G}\cdot {\bf r}_{ij}\right) \;,
  \end{split}
\end{equation}
where $\Theta$ is the Heaviside step function, $\{{\bf G} \in s{\rm
th~star}\}$ are the reciprocal lattice vectors of the simulation cell
in the $s$th star of symmetry-equivalent vectors, and $\{\alpha_l\}$,
$\{a_s\}$ and $L_u$ are optimizable parameters.
Note that we impose the parallel-spin Kato cusp condition
\cite{suppl_kato_1957} by setting $\alpha_0 = (\alpha_1 - 1/4) L_u/3$.

In our multideterminantal benchmark of FCIQMC we replace the HF
determinant with a selected-CI expansion extracted from FCIQMC.

Backflow transformations \cite{suppl_pandharipande_backflow_1973,
suppl_schmidt_backflow_1979, suppl_kwon_backflow_1998,
suppl_holzmann_backflow_2003, suppl_lopezrios_backflow_2006} offer the
ability to modify the nodes of the Slater-Jastrow wave function and
give significantly lower DMC energies.
Backflow transformations replace the argument $\bf R$ of the Slater
determinants with transformed coordinates ${\bf X}({\bf R})$ which
we parametrize as \cite{suppl_lopezrios_backflow_2006}
\begin{equation}
  \label{Seq:backflow}
  {\bf x}_i = {\bf r}_i + \sum_{j\neq i}
              \left(1-r_{ij}/L_\eta\right)^3
              \Theta\left(r_{ij}-L_\eta\right)
              \sum_{l=0}^8 c_l r_{ij}^l {\bf r}_{ij} \;,
\end{equation}
where $\{c_l\}$ and $L_\eta$ are optimizable parameters.
Note that we set $c_0 = c_1 L_\eta/3$ to avoid interfering with the
parallel-spin Kato cusp condition.

\section{Full configuration-interaction quantum Monte Carlo}

The FCIQMC method \cite{suppl_booth_fciqmc_2009,
suppl_cleland_initiator_2010, suppl_cleland_affinities_2011,
suppl_booth_c2_2011} obtains the CI coefficients by evolution of a
population of random walkers, each associated with a determinant in
Hilbert space, in imaginary time via diagonal death/cloning and
off-diagonal spawning processes.
An annihilation step is carried out at each time step to cancel
walkers of opposite signs on the same determinant, which is crucial
for sign coherence \cite{suppl_booth_fciqmc_2009}.
The initial set of walkers, $100$ in our calculations, is usually
placed on the HF determinant, and after an equilibration stage the
occupation of each determinant is on average proportional to its exact
CI coefficient.

The initiator approximation modifies the dynamics of the random walk
so that spawning new walkers on unpopulated determinants from sites
that contain less than $n_{\rm init}$ walkers is forbidden, where
$n_{\rm init}$ is a tunable parameter which we set to 3 in our
calculations.
This allows a substantial reduction in the number of walkers $W$
required for convergence, but is a source of bias
\cite{suppl_cleland_initiator_2010, suppl_cleland_affinities_2011,
suppl_booth_c2_2011}.
The initiator error vanishes as $W\to\infty$, and in practice we
increase the walker population until energy changes become negligible.
Our largest calculations use up to $W=1.5 \times 10^8$ walkers.
The number of walkers required to overcome the initiator error
increases with the size of the Hilbert space of the system, which
grows very quickly with system size, and has also been observed to
increase with $r_{\rm s}$ \cite{suppl_shepherd_fciqmc_2012,
suppl_shepherd_ueg_2012}.

\begin{figure}[!hbt]
  \begin{center}
    \includegraphics[width=\columnwidth]{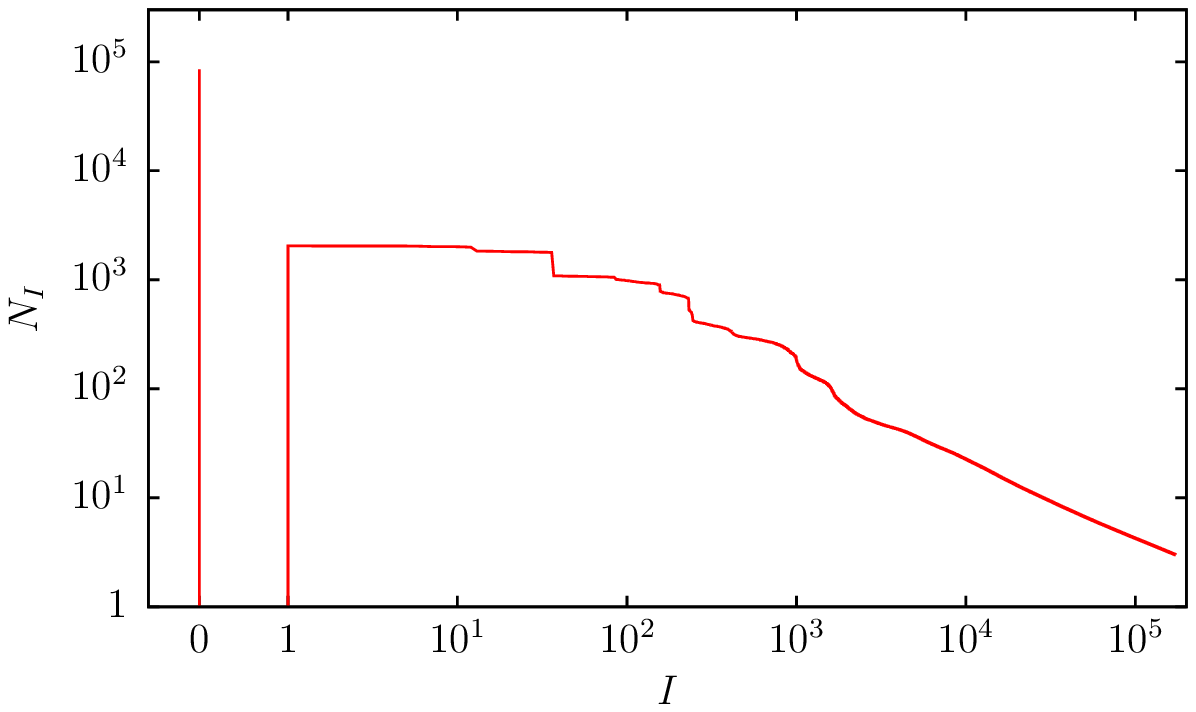}
    \caption{
      Walker population $N_I$ on Slater determinant $D_I$ as a
      function of $I$, sorted by decreasing $N_I$, for the
      $19$-electron gas at $r_{\rm s} =1$ and ${\bf k}_{\rm s} =
      {\bf 0}$ using a 341-plane-wave basis and $10^7$ walkers.
      The first peak corresponds to the HF determinant.}
    \label{Sfig:wf}
  \end{center}
\end{figure}

Figure \ref{Sfig:wf} represents the equilibrated walker population on
the leading determinants of the CI wave function for one of the
systems reported in our work.

\subsection{Basis-set extrapolation}

Basis sets for the UEG consist of the $M$ plane waves with the
smallest wave vectors.
This finite basis set provides access to a finite portion of the
Hilbert space of the system, resulting in a positive energy bias.
The infinite basis set limit can be estimated by extrapolation, as is
standard practice in quantum chemistry
\cite{suppl_helgaker_mol_es_2014}.
We extrapolate our FCIQMC correlation energies at each IBZ region $z$
to the complete basis set limit using the fitting function
\begin{equation}
  \label{Seq:basis-set-extrapolation}
  E_{\rm corr}^z(M) = E_{\rm corr}^z(\infty)
    + a_z M^{-1} + b_z M^{-2} \;,
\end{equation}
where $E_{\rm corr}^z(\infty)$, $a_z$, and $b_z$ are fit parameters.
Setting $a_z = 0$ yields a very good fit to the energy data, but we
keep $a_z$ as a fit parameter to account for the parametrization bias.

\begin{figure}[!hbt]
  \begin{center}
    \includegraphics[width=\columnwidth]{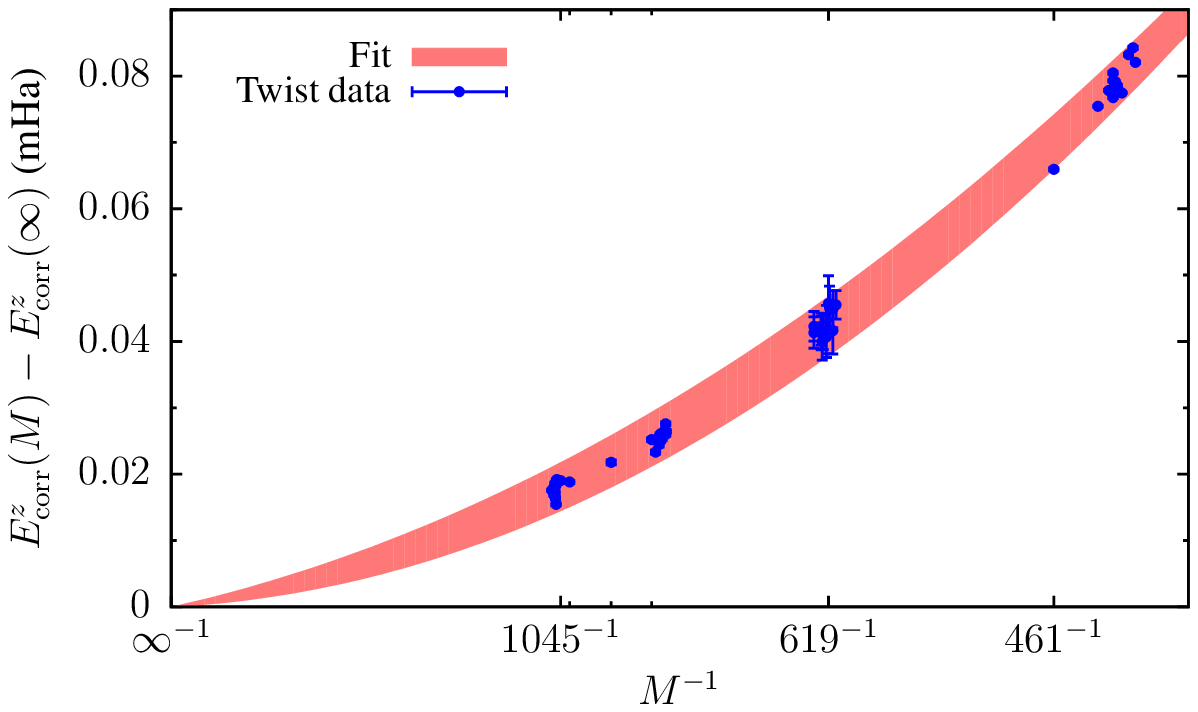}
    \caption{
      Finite basis set error in the FCIQMC correlation energy as a
      function of inverse basis-set size $M^{-1}$ for the
      $19$-electron gas at $r_{\rm s} = 1$.
      The data correspond to calculations at multiple basis-set sizes
      in each of the 13 IBZ regions.}
    \label{Sfig:conv}
  \end{center}
\end{figure}

We find that the basis-set error is roughly independent of $z$, as
shown in Fig.\ \ref{Sfig:conv} for the 19-electron gas at $r_{\rm
s}=1$.
To reduce the cost of our FCIQMC calculations for selected systems
($N=19$, $27$, and $33$ at $r_{\rm s}=0.5$) we perform the basis set
extrapolation in the $\Gamma$-point region, $z=0$, only.
For other regions we evaluate the correlation energy at a single basis
set size $M$ ($\simeq 1050$, $830$, and $830$, respectively) and we
obtain $E_{\rm corr}^z(\infty)$ from Eq.\
\ref{Seq:basis-set-extrapolation} by setting $a_z=a_0$ and $b_z=b_0$.

We investigate the bias incurred by this approximation by comparing
the value of $b_0$ with its average $b_{\rm ave} = \sum_z
\left(\Omega_z/\Omega_{\rm BZ}\right) b_z$ in fits of our data at
$r_{\rm s}=1$ to Eq.\ \ref{Seq:basis-set-extrapolation} with $a_z=0$.
The largest deviation occurs for the $27$-electron gas, for which
$\vert 1 - b_0/b_{\rm ave}\vert = 0.153$.
Therefore a contribution of $0.153 \times b_0 M^{-2}$ is added (in
quadrature) to the uncertainty of the twist-averaged energy for the
systems for which we use the $\Gamma$-point extrapolation scheme,
which we expect to overestimate the corresponding bias.
This correction represents an increase in the uncertainty of the
twist-averaged correlation energy by up to $75\%$, but despite this,
the $\Gamma$-point extrapolation method provides a net reduction in
the computational cost of the FCIQMC calculations.

\section{Finite size errors}

\subsection{Finite system results and extrapolation}

\begin{figure}[!hbt]
  \begin{center}
    \includegraphics[width=\columnwidth]{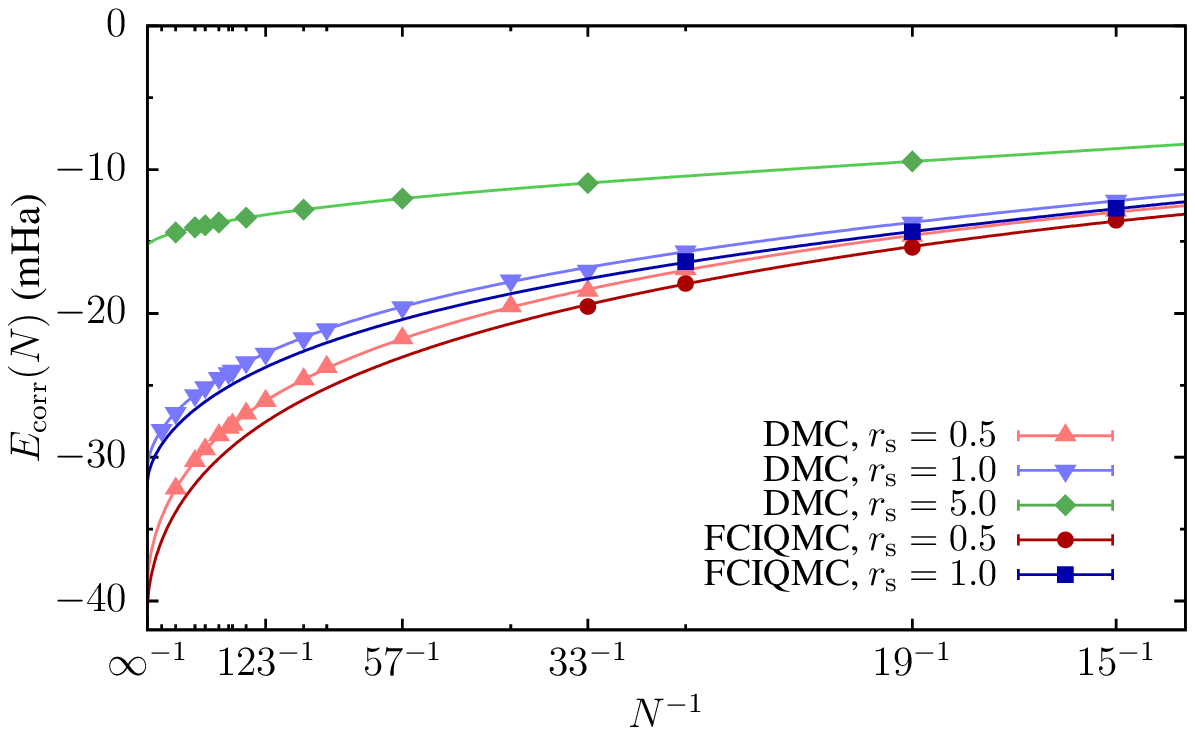}
    \caption{
      Full set of twist-averaged DMC and FCIQMC correlation energies
      as a function of inverse system size, along with finite-size
      extrapolation fits described in the manuscript.
      The uncertainty in the fitted curves is smaller than the line
      width.}
    \label{Sfig:eos}
  \end{center}
\end{figure}

In Table \ref{Stab:dmc} we give the full set of twist-averaged DMC
correlation energies obtained in our work, and Table \ref{Stab:fciqmc}
shows the FCIQMC results.
For completeness, we also give total-energy versions of these data
in Tables \ref{Stab:dmc_etot} and \ref{Stab:fciqmc_etot}.
We plot our DMC and FCIQMC correlation energies in Fig.\
\ref{Sfig:eos}, along with the corresponding fits.

\begin{table}[!hbt]
  \begin{center}
    \begin{tabular}{r@{\quad}r@{.}l@{\quad}r@{.}l@{\quad}r@{.}l}
      \hline
      \multicolumn{1}{r}{} &
      \multicolumn{6}{c}{$E_{\rm corr}^{\rm FN}$} \\
      \multicolumn{1}{c}{$N$}             &
      \multicolumn{2}{c}{$r_{\rm s}=0.5$} &
      \multicolumn{2}{c}{$r_{\rm s}=1.0$} &
      \multicolumn{2}{c}{$r_{\rm s}=5.0$} \\
      \hline
      \hline
      $  15$ & $-12$&$883(6) $ & $-12$&$144(3) $ &
                                             \multicolumn{2}{c}{} \\
      $  19$ & $-14$&$636(10)$ & $-13$&$665(8) $ & $ -9$&$425(6)$ \\
      $  27$ & $-16$&$953(9) $ & $-15$&$654(9) $ &
                                             \multicolumn{2}{c}{} \\
      $  33$ & $-18$&$455(9) $ & $-17$&$009(34)$ & $-10$&$943(6)$ \\
      $  40$ & $-19$&$526(94)$ & $-17$&$683(30)$ &
                                             \multicolumn{2}{c}{} \\
      $  57$ & $-21$&$757(64)$ & $-19$&$549(47)$ & $-12$&$017(6)$ \\
      $  81$ & $-23$&$750(36)$ & $-21$&$065(22)$ &
                                             \multicolumn{2}{c}{} \\
      $  93$ & $-24$&$598(9) $ & $-21$&$709(14)$ & $-12$&$779(4)$ \\
      $ 123$ & $-26$&$108(33)$ & $-22$&$769(17)$ &
                                             \multicolumn{2}{c}{} \\
      $ 147$ & $-26$&$984(20)$ & $-23$&$373(33)$ & $-13$&$344(4)$ \\
      $ 171$ & $-27$&$762(11)$ & $-23$&$956(7) $ &
                                             \multicolumn{2}{c}{} \\
      $ 179$ & $-27$&$981(12)$ & $-24$&$135(15)$ &
                                             \multicolumn{2}{c}{} \\
      $ 203$ & $-28$&$493(15)$ & $-24$&$463(11)$ & $-13$&$677(3)$ \\
      $ 251$ & $-29$&$487(14)$ & $-25$&$099(18)$ & $-13$&$872(6)$ \\
      $ 305$ & $-30$&$294(13)$ & $-25$&$660(7) $ & $-14$&$021(2)$ \\
      $ 515$ & $-32$&$204(6) $ & $-26$&$892(2) $ & $-14$&$3616(5)$\\
      $1021$ & \multicolumn{2}{c}{}
                               & $-28$&$085(3) $ &
                                   \multicolumn{2}{c}{} \\[0.1cm]
      $\infty$
             & $-38$&$778(10)$ & $-30$&$650(3) $ & $-15$&$270(4)$ \\
      \hline
    \end{tabular}
  \end{center}
  \caption{Twist-averaged DMC correlation energies of the polarized
    UEG at $r_{\rm s} = 0.5$, $1$, and $5$ and their respective
    thermodynamic limits, in mHa.}
  \label{Stab:dmc}
\end{table}

\begin{table}[!hbt]
  \begin{center}
    \begin{tabular}{r@{\quad}r@{.}lr@{.}l@{\quad}r@{.}lr@{.}l}
      \hline
      \multicolumn{1}{r}{} &
      \multicolumn{4}{c}{$r_{\rm s}=0.5$} &
      \multicolumn{4}{c}{$r_{\rm s}=1.0$} \\
      \multicolumn{1}{c}{$N$}                     &
      \multicolumn{2}{c}{$E_{\rm corr}$}          &
      \multicolumn{2}{c}{$\varepsilon_{\rm FN}$}  &
      \multicolumn{2}{c}{$E_{\rm corr}$}          &
      \multicolumn{2}{c}{$\varepsilon_{\rm FN}$}  \\
      \hline
      \hline
      $15$ & $-13$&$5203(7) $ & $  0$&$638(6) $
           & $-12$&$6926(4) $ & $  0$&$549(3) $ \\
      $19$ & $-15$&$404(3)  $ & $  0$&$769(10)$
           & $-14$&$313(4)  $ & $  0$&$648(9) $ \\
      $27$ & $-17$&$918(7)  $ & $  0$&$965(12)$
           & $-16$&$395(1)  $ & $  0$&$741(9) $ \\
      $33$ & $-19$&$516(13) $ & $  1$&$061(15)$
           & \multicolumn{4}{c}{} \\[0.1cm]
      $\infty$ &
             $-40$&$44(5)   $ & $  1$&$67(5)  $
           & $-31$&$70(4)   $ & $  1$&$05(4)  $ \\
      \hline
    \end{tabular}
  \end{center}
  \caption{
    Exact (FCIQMC) correlation energies and fixed node error for the
    polarized UEG at $r_{\rm s} = 0.5$ and $1$ for different system
    sizes and their respective thermodynamic limits, in mHa.}
  \label{Stab:fciqmc}
\end{table}

\begin{table}[!hbt]
  \begin{center}
    \begin{tabular}{r@{\quad}l@{\quad}l@{\quad}l}
      \hline
      \multicolumn{1}{r}{} &
      \multicolumn{3}{c}{$E_{\rm tot}^{\rm FN}$} \\
      \multicolumn{1}{c}{$N$} &
      \multicolumn{1}{c}{$r_{\rm s}=0.5$} &
      \multicolumn{1}{c}{$r_{\rm s}=1.0$} &
      \multicolumn{1}{c}{$r_{\rm s}=5.0$} \\
      \hline
      \hline
      $  15$ & $5.763978(6) $ & $1.116571(3) $ &                \\
      $  19$ & $5.772742(10)$ & $1.121587(8) $ & $-0.063725(6)$ \\
      $  27$ & $5.786617(9) $ & $1.128388(9) $ &                \\
      $  33$ & $5.797524(86)$ & $1.132718(34)$ & $-0.062320(6)$ \\
      $  40$ & $5.794355(94)$ & $1.133105(30)$ &                \\
      $  57$ & $5.801068(64)$ & $1.136407(47)$ & $-0.061699(6)$ \\
      $  81$ & $5.804203(36)$ & $1.138375(22)$ &                \\
      $  93$ & $5.807672(9) $ & $1.139660(14)$ & $-0.061268(4)$ \\
      $ 123$ & $5.810636(33)$ & $1.141072(17)$ &                \\
      $ 147$ & $5.811364(20)$ & $1.141570(33)$ & $-0.061032(4)$ \\
      $ 171$ & $5.814004(12)$ & $1.142477(7) $ &                \\
      $ 179$ & $5.814569(12)$ & $1.142664(15)$ &                \\
      $ 203$ & $5.814274(15)$ & $1.142746(11)$ & $-0.060903(3)$ \\
      $ 251$ & $5.816088(14)$ & $1.143470(18)$ & $-0.060833(6)$ \\
      $ 305$ & $5.816817(13)$ & $1.143789(7) $ & $-0.060788(2)$ \\
      $ 515$ & $5.819314(6) $ & $1.144758(2) $ & $-0.0606891(5)$\\
      $1021$ &                & $1.145350(3) $ &
                                                \\[0.1cm]
      $\infty$
             & $5.822717(10)$ & $1.146098(3) $ & $-0.060560(4)$ \\
      \hline
    \end{tabular}
  \end{center}
  \caption{Twist-averaged DMC total energies of the polarized
    UEG at $r_{\rm s} = 0.5$, $1$, and $5$ and their respective
    thermodynamic limits, in Ha.}
  \label{Stab:dmc_etot}
\end{table}

\begin{table}[!hbt]
  \begin{center}
    \begin{tabular}{r@{\quad}l@{\quad}l}
      \hline
      \multicolumn{1}{r}{}                &
      \multicolumn{2}{c}{$E_{\rm tot}$}   \\
      \multicolumn{1}{c}{$N$}             &
      \multicolumn{1}{c}{$r_{\rm s}=0.5$} &
      \multicolumn{1}{c}{$r_{\rm s}=1.0$} \\
      \hline
      \hline
      $15$ & $5.7633402(7)  $ & $1.1160227(4) $ \\
      $19$ & $5.7719736(26) $ & $1.1209393(39)$ \\
      $27$ & $5.7856521(72) $ & $1.1276472(14)$ \\
      $33$ & $5.7964623(129)$ &                 \\[0.1cm]
      $\infty$ &
             $5.82105(5)    $ & $1.14505(4)   $ \\
      \hline
    \end{tabular}
  \end{center}
  \caption{
    Exact (FCIQMC) total energies for the polarized UEG at
    $r_{\rm s} = 0.5$ and $1$ for different system sizes and their
    respective thermodynamic limits, in Ha.}
  \label{Stab:fciqmc_etot}
\end{table}

\begin{figure}[!hbt]
  \begin{center}
    \includegraphics[width=\columnwidth]{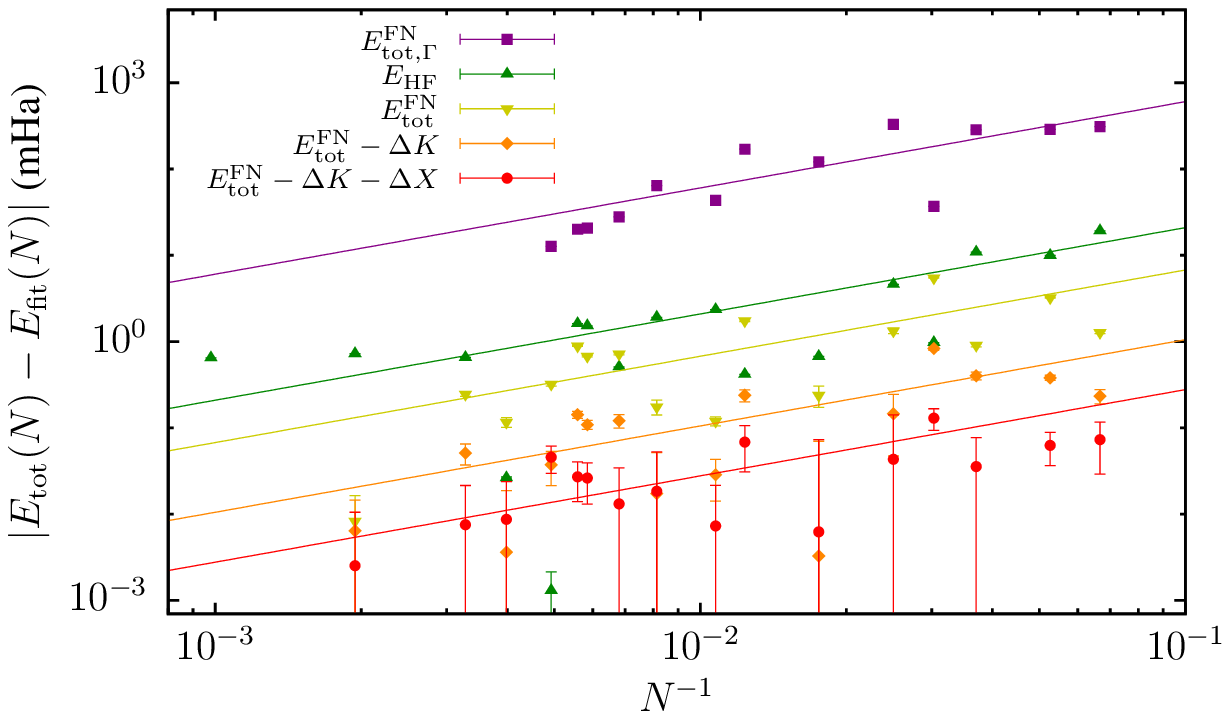}
    \caption{
      Absolute fit error as a function of $N^{-1}$ for the HF and
      fixed node DMC energy at $r_{\rm s}=0.5$ with various
      corrections.
      The lines are functions proportional to $N^{-1}$, intended as a
      guide to the eye.}
    \label{Sfig:sjdmc_qrandom}
  \end{center}
\end{figure}

To illustrate the magnitude of the quasirandom fluctuations in the DMC
energy as a function of $N$, in Fig.\ \ref{Sfig:sjdmc_qrandom} we plot
the fit error for the $\Gamma$-point DMC energy, the twist-averaged HF
energy, and the twist-averaged DMC energy with and without the $\Delta
K$ and $\Delta X$ corrections at $r_{\rm s}=0.5$.
For the $\Gamma$-point DMC energy we have used a fitting function
consisting of a single $N^{-1}$ term, while in the other cases we have
used fitting functions containing up to $N^{-2}$, as in Eq.\ 1 of our
manuscript.
These fit errors are expected to be proportional to $N^{-1}$, and in
Fig.\ \ref{Sfig:sjdmc_qrandom} we plot straight lines of this form to
guide the eye.

At this density, twist-averaging reduces the magnitude of the
quasirandom fluctuations by two orders of magnitude.
The $\Delta K$ correction affords a further reduction by a factor of
$7$, while $\Delta X$ achieves an additional factor of $4$.
We note that the fit errors in $E_{\rm tot}^{\rm FN}-\Delta K-\Delta
X$ are of the order of the uncertainty to which we have evaluated the
DMC energies.

\subsection{Evaluation of integration errors at $\kvec={\bf 0}$}

As shown in Refs.\ \onlinecite{suppl_chiesa_fse_2006,
suppl_drummond_fse_2008}, some of the leading-order finite-size errors
in DMC energies can be ascribed to integration errors which are
effectively due to the inability to sample $\kvec={\bf 0}$ at finite
$N$.
We define the object
\begin{equation}
  \label{Seq:epsilon_def}
  \begin{split}
    \epsilon_n(\alpha)
      = \Omega^{\frac{n+1}{3}} {\Bigg [} &
          (2\pi)^{-3}
          \int \frac{4\pi}{k^2} k^n e^{-\alpha k^2} \,{\rm d}\kvec \\
      & - \frac 1 \Omega
          \sum_{{\bf G}\neq {\bf 0}} \frac{4\pi}{G^2}
             G^n e^{-\alpha G^2} {\Bigg ]} \;,
  \end{split}
\end{equation}
where $\Omega$ is the simulation cell volume, whose limit $\epsilon_n
= \lim_{\alpha\to 0} \epsilon_n(\alpha)$ represents the error in the
discretization of the reciprocal-space convolution of the interaction
potential $\frac{4\pi}{k^2}$ and a power of the wave vector $k^n$.

The finite size error in the HF exchange energy of the polarized UEG
due to integrations errors at $\kvec={\bf 0}$ (given in Eq.\ 41 of
Ref.\ \onlinecite{suppl_drummond_fse_2008} for the unpolarized UEG)
can be written in terms of $\epsilon_n$ as
\begin{equation}
  \label{Seq:X_fse}
  \begin{split}
  X(N) & = X(\infty)
         - \frac{3 \epsilon_1}{16\pi} r_{\rm s}^{-1} N^{-2/3} \\
       & + \frac{\epsilon_3}{(4\pi)^3} \left(\frac \pi 6\right)^{2/3}
           r_{\rm s}^{-1} N^{-4/3}
         + \ldots \;.
  \end{split}
\end{equation}
Note that, in the notation of Ref.\
\onlinecite{suppl_drummond_fse_2008}, $\epsilon_1 = 2 C_{\rm HF}$.
The finite size error in the DMC kinetic energy of the polarized UEG
due to integration errors at $\kvec={\bf 0}$ (given in Eq.\ 56 of
Ref.\ \onlinecite{suppl_drummond_fse_2008}) can also be expressed in
terms of $\epsilon_n$,
\begin{equation}
  \label{Seq:T_fse}
  \begin{split}
  T(N) & = T(\infty)
         - \frac{\sqrt{3}} 4 r_{\rm s}^{-3/2} N^{-1} \\
       & + \frac{\epsilon_3}{16\pi} r_{\rm s}^{-2} N^{-4/3}
         + \ldots \;.
  \end{split}
\end{equation}
Note that, in the notation of Ref.\
\onlinecite{suppl_drummond_fse_2008}, $\epsilon_3 = 4 C_{\rm 3D}$.

By manipulating Eq.\ \ref{Seq:epsilon_def} we arrive at a computable
expression for $\epsilon_n(\alpha)$,
\begin{equation}
  \label{Seq:epsilon}
  \epsilon_n(\alpha)
    = \Omega^{\frac{n+1}{3}} \left[
        \frac {\Gamma\left(\frac{n+1} 2\right)}
              {\pi \alpha^{\frac{n+1} 2}}
      - \frac{4\pi}\Omega
        \sum_{{\bf G}\neq {\bf 0}} G^{n-2} e^{-\alpha G^2} \right] \;,
\end{equation}
where $\Gamma$ is the Gamma function.
The numerical evaluation of $\epsilon_n$ requires computing
$\epsilon_n(\alpha)$ at increasingly small values of $\alpha$ until a
convergence criterion is met.
As can be gathered from Eq.\ \ref{Seq:epsilon}, $\epsilon_n(\alpha)$
at $\alpha\to 0$ is the difference of increasingly large numbers, one
of which is itself an infinite sum which needs to be converged
independently.
This is numerically delicate, and we find that rounding errors prevent
obtaining more than 4--5 decimal places of precision in the value of
$\epsilon_n$ with this procedure.

\begin{figure}[!hbt]
  \begin{center}
    \includegraphics[width=\columnwidth]{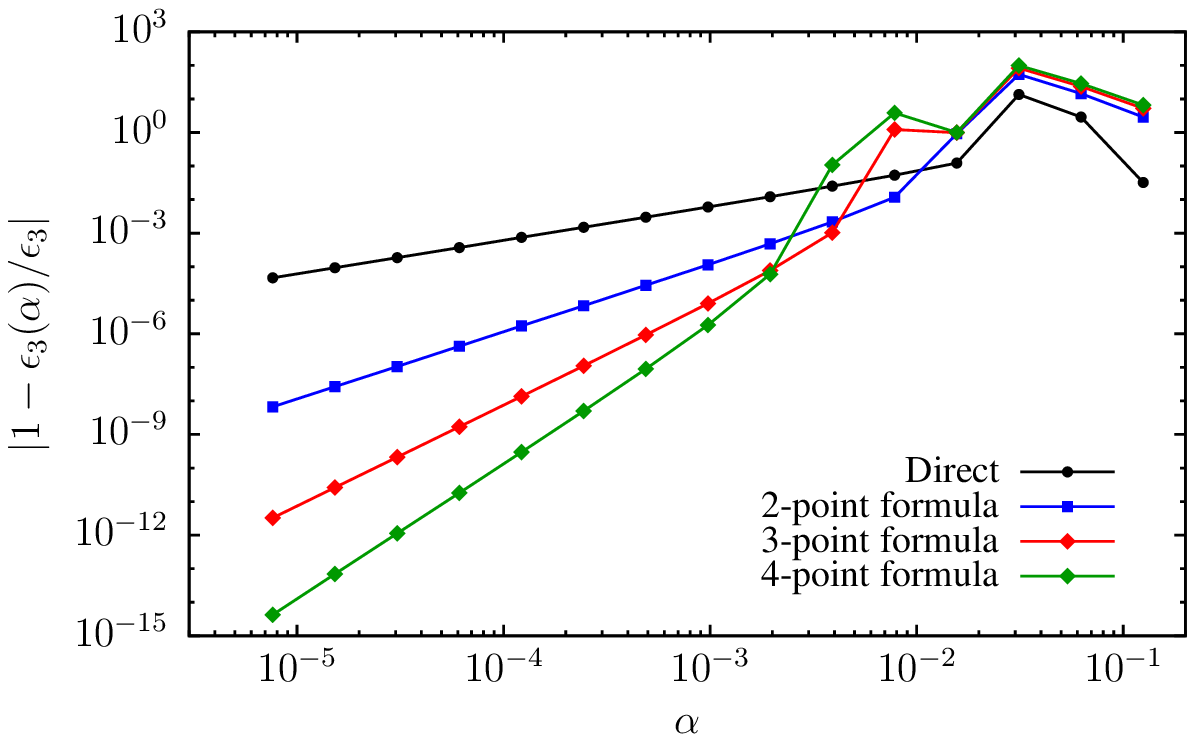}
    \caption{
      Convergence of the integration error $\epsilon_3(\alpha)$ as a
      function of $\alpha$, along with extrapolated estimates from the
      two-point formula of Eq.\ \ref{Seq:epsilon_2point} and the
      three-point formula of Eq.\ \ref{Seq:epsilon_3point}.}
    \label{Sfig:epsilon_accel}
  \end{center}
\end{figure}

However, inspection of the behavior of $\epsilon_n(\alpha)$ with
$\alpha$ reveals an exponential convergence pattern, which can be
exploited to produce much more accurate estimates of $\epsilon_n$ at
values of $\alpha$ at which rounding errors are not problematic.
Using the model $\epsilon_n(\alpha) = \epsilon_n e^{-p_1 \alpha}$ we
find a two-point extrapolation formula,
\begin{equation}
  \label{Seq:epsilon_2point}
  \epsilon_n \approx \epsilon_n^2(\alpha)
                     \epsilon_n^{-1}(2\alpha) \;,
\end{equation}
a higher-order model $\epsilon_n(\alpha) = \epsilon_n e^{-p_1
\alpha -p_2 \alpha^2}$ yields a three-point extrapolation formula,
\begin{equation}
  \label{Seq:epsilon_3point}
  \epsilon_n \approx \epsilon_n^{8/3}(\alpha)
                     \epsilon_n^{-2}(2\alpha)
                     \epsilon_n^{1/3}(4\alpha) \;,
\end{equation}
and a three-parameter model $\epsilon_n(\alpha) = \epsilon_n e^{-p_1
\alpha -p_2 \alpha^2 -p_3 \alpha^3}$ results in a four-point
extrapolation formula,
\begin{equation}
  \label{Seq:epsilon_4point}
  \epsilon_n \approx \epsilon_n^{64/21}(\alpha)
                     \epsilon_n^{-8/3}(2\alpha)
                     \epsilon_n^{2/3}(4\alpha)
                     \epsilon_n^{-1/21}(8\alpha) \;,
\end{equation}
We plot the values of $\epsilon_3(\alpha)$ and the results from the
three extrapolation formulae in Fig.\ \ref{Sfig:epsilon_accel}.
This technique significantly accelerates convergence:\@ Eq.\
\ref{Seq:epsilon_4point} gives $\epsilon_3$ to 14-digit precision at a
value of $\alpha$ at which $\epsilon_3(\alpha)$ is only accurate to 4
decimal places.
We note that we have used 128-bit floating-point arithmetic (``quad''
precision) to further enhance numerics.
With this approach we obtain the values $\epsilon_1=5.674594959$ and
$\epsilon_3=21.04959845$ for our simple cubic simulation cell.

\subsection{Other sources of finite-size errors}

Besides integration errors and quasirandom fluctuations, there is a
third source of finite-size errors in twist-averaged energies.
As reported in Table I of Ref.\ \onlinecite{suppl_lin_ta_2001}, the
twist-averaged HF kinetic energy exhibits finite-size errors that
scale as $N^{-4/3}$ to leading order.
These finite size errors arise due to the use of the canonical
ensemble, \textit{i.e.}, keeping $N$ fixed as $\kvec_{\rm s}$ is
varied during twist-averaging, and is associated with the mismatch
between the Fermi wave vector at size $N$ and in the thermodynamic
limit \cite{suppl_lin_ta_2001}.
In other words, these are integration errors at $k=k_{\rm
F}$ due to the smearing of the Fermi surface as an artifact of
twist-averaging in the canonical ensemble.

We observe in our data that the estimate of the finite-size error at
order $N^{-4/3}$ in the HF exchange energy given by Eq.\
\ref{Seq:X_fse} and in the DMC kinetic energy given by Eq.\
\ref{Seq:T_fse} do not completely account for the finite-size error at
order $N^{-4/3}$ in either of these energy components or in the DMC
correlation energy.
We hypothesize that integration errors at $k=k_{\rm F}$ from the
various energy components enter the DMC correlation energy at order
$N^{-4/3}$ and would need to be fully accounted for in order to
determine the $c_4$ coefficient in Eq.\ 1 of our manuscript.
For this reason we treat all coefficients beyond order $N^{-1}$ as fit
parameters in our analysis of the correlation energies.

Our DMC correlation energies are well described by Eq.\ 1 of our
manuscript with $c_6=0$, but we keep $c_6$ as a fit parameter to
account for the parametrization error.

\section{Additional results}

\subsection{Accuracy of the $\xi$-scale extrapolation}

In our manuscript we find that correlation energies at high densities
are accurately described by polynomials in $\xi=r_{\rm s}^{-3/2}
N^{-1}$ with density-independent coefficients.
The fit shown in Fig.\ 4 of our manuscript yields fixed node
correlation energies in the thermodynamic limit of $-38.722(8)$ and
$-30.658(2)$ mHa at $r_{\rm s}=0.5$ and 1, respectively, which differ
by $4.3$ and $2.0$ standard deviations from the values obtained by
independent extrapolation at each density given in Table
\ref{Stab:dmc}.

\subsection{Modified PW92 parameters}

In our manuscript we present values of the correlation energy from
alternative fits to the PW92 form for the correlation energy of the
polarized UEG \cite{suppl_perdew_wang_1992}.
In particular, we introduce an unweighted fit to the CA data
\cite{suppl_ceperley_1980} (uPW92) and a revised unweighted fit to the
CA data and our results (rPW92).
In Table \ref{Stab:pw_params} we provide the parameter values that
reproduce the mean values of the various fits.

\begin{table}[!hbt]
  \begin{center}
    \begin{tabular}{l@{\quad}lll}
      \hline
                 &
      \multicolumn{1}{c}{$\alpha_1$} &
      \multicolumn{1}{c}{$\beta_3$}  &
      \multicolumn{1}{c}{$\beta_4$}  \\
      \hline
      \hline
      PW92     & $0.20548 $ & $3.3662 $ & $0.62317 $ \\
      PW92$^*$ & $0.202326$ & $3.30573$ & $0.615932$ \\
      uPW92    & $0.264193$ & $4.78287$ & $0.750424$ \\
      rPW92    & $0.266529$ & $4.86059$ & $0.750188$ \\
      \hline
    \end{tabular}
  \end{center}
  \caption{Values of the free parameters in the PW92 parametrization
    of the correlation energy of the polarized UEG, named $\alpha_1$,
    $\beta_3$, and $\beta_4$ following the notation of Ref.\
    \onlinecite{suppl_perdew_wang_1992}.
    Listed are the values given in Ref.\
    \onlinecite{suppl_perdew_wang_1992} (PW92), the similar values
    we obtain with a weighted fit to the CA data using our fitting
    methodology (PW92$^*$), the values we obtain with an unweighted
    fit to the CA data (uPW92), and the values we obtain with an
    unweighted fit to the CA data and our present results (rPW92).}
  \label{Stab:pw_params}
\end{table}

The parameters in the uPW92 and rPW92 fits have very similar values,
and, while this might well be coincidental, it is of potential
interest to report the equivalent uPW92 fit for the unpolarized UEG,
which we have not considered in our work.
In Table \ref{Stab:pw_params_unpol} we provide the parameter values
that reproduce the mean values of the unweighted PW92 fit to the CA
data for the unpolarized UEG.

\begin{table}[!hbt]
  \begin{center}
    \begin{tabular}{l@{\quad}lll}
      \hline
                 &
      \multicolumn{1}{c}{$\alpha_1$} &
      \multicolumn{1}{c}{$\beta_3$}  &
      \multicolumn{1}{c}{$\beta_4$}  \\
      \hline
      \hline
      PW92     & $0.21370 $ & $1.6382 $ & $0.49294 $ \\
      PW92$^*$ & $0.216518$ & $1.66722$ & $0.498875$ \\
      uPW92    & $0.227012$ & $1.76522$ & $0.523918$ \\
      \hline
    \end{tabular}
  \end{center}
  \caption{Values of the free parameters in the PW92 parametrization
    of the correlation energy of the unpolarized UEG, named
    $\alpha_1$, $\beta_3$, and $\beta_4$ following the notation of
    Ref.\ \onlinecite{suppl_perdew_wang_1992}.
    Listed are the values given in Ref.\
    \onlinecite{suppl_perdew_wang_1992} (PW92), the similar values
    we obtain with a weighted fit to the CA data using our fitting
    methodology (PW92$^*$), and the values we obtain with an
    unweighted fit to the CA data (uPW92).}
  \label{Stab:pw_params_unpol}
\end{table}

\FloatBarrier

\end{document}